\documentclass[aps,pre,twocolumn,showpacs,superscriptaddress,longbibliography,nofootinbib]{revtex4-2}

\usepackage{amsmath,amssymb}
\usepackage{graphicx}
\usepackage{bm}
\usepackage{xcolor}
\usepackage{hyperref}
\usepackage{siunitx}
\usepackage{booktabs}

\newcommand{\note}[1]{\textcolor{black}{#1}}

\begin{document}

% ============================================================
%  TITLE BLOCK
% ============================================================
\title{Morphology-Resolved Stress Contributions in Sheared Wet Granular Materials}

\author{Ahmad Awdi}
\affiliation{Laboratoire NAVIER, Universit\'e Gustave Eiffel, ENPC,
Institut Polytechnique de Paris, CNRS, 77420 Champs-sur-Marne}

\author{Camille Chateau}
\affiliation{Laboratoire NAVIER, Universit\'e Gustave Eiffel, ENPC,
Institut Polytechnique de Paris, CNRS, 77420 Champs-sur-Marne}

\author{Coumba Niang}
\affiliation{Laboratoire NAVIER, Universit\'e Gustave Eiffel, ENPC,
Institut Polytechnique de Paris, CNRS, 77420 Champs-sur-Marne}

\author{Patrick Aimedieu}
\affiliation{Laboratoire NAVIER, Universit\'e Gustave Eiffel, ENPC,
Institut Polytechnique de Paris, CNRS, 77420 Champs-sur-Marne}

\author{Jean-No\"el Roux}
\affiliation{Laboratoire NAVIER, Universit\'e Gustave Eiffel, ENPC,
Institut Polytechnique de Paris, CNRS, 77420 Champs-sur-Marne}

\author{Abdoulaye Fall}
\email{abdoulaye.fall@cnrs.fr}
\affiliation{Laboratoire NAVIER, Universit\'e Gustave Eiffel, ENPC,
Institut Polytechnique de Paris, CNRS, 77420 Champs-sur-Marne}

\date{\today}

% ============================================================
%  ABSTRACT
% ============================================================
\begin{abstract}
\note{Three-dimensional X-ray microtomography, coupled to rheometric measurements, enables a morphology-resolved reconstruction of capillary stresses at the grain scale in unsaturated wet granular materials. Liquid domains are automatically classified into capillary bridges, dimers, trimers, and larger clusters, and their spatial organization is tracked as a function of shear deformation and liquid content. We show that shear localization governs the redistribution of the liquid phase: capillary bridges remain uniformly distributed throughout the sample, while higher-order morphologies accumulate preferentially near the lower boundary of the shear-zone through a shear-driven coalescence mechanism. Despite this spatial localization, simple two-grain bridges generate the dominant contribution to the isotropic capillary pressure, accounting for nearly 85\% of the total at liquid-to-solid volume ratio $\epsilon = 0.05$, whereas more complex liquid clusters contribute only weakly to the overall cohesion. Incorporating the morphology-resolved capillary pressure into an effective-stress framework qualitatively reproduces the macroscopic friction coefficient across the full range of investigated liquid contents, without adjustable parameters. These results establish a predictive micro--macro link between liquid morphology and the rheology of wet granular materials.}
\end{abstract}


\maketitle

% ============================================================
%  SECTION I — INTRODUCTION
% ============================================================
\section{Introduction}
\label{sec:intro}

\note{Wetting a granular material with a small amount of liquid profoundly alters its mechanical response. Even minute liquid fractions generate capillary bridges between neighboring grains, inducing cohesion, modifying shear localization, and increasing macroscopic strength~\cite{Mitarai2006,Mani2013,Singh2014,Khamseh2015,Berger2016,Vo2020,Awdi2023,Badetti2018a,Badetti2018b,Amarsid2024}. This phenomenon underlies a wide range of systems, from geophysical flows and landslides to powder processing, food technology, and additive manufacturing. Despite its ubiquity, the micro-to-macro connection between liquid morphology and bulk rheology remains incompletely understood.}

\note{The mechanics of individual capillary bridges are now well characterized. Analytical and
numerical studies have established the dependence of capillary forces on bridge volume,
separation distance, and contact angle, while discrete-element and lattice-Boltzmann simulations have extended this understanding to small clusters and funicular
regimes~\cite{Willett2000,Huang2015,Lian1993,Delenne2015,Younes2023,Richefeu2016}. At the
macroscopic scale, cohesive granular materials exhibit modified yield criteria, enhanced apparent friction, and departures from the dry $\mu(I)$ rheology~\cite{Mani2013,Singh2014,Khamseh2015,Berger2016,Vo2020,Awdi2023,Badetti2018a,Mandal2020}. Quantitative models are most often restricted to the pendular regime, where liquid exists primarily as isolated capillary bridges between grains.} In this limit, wet granular materials behave similarly to other cohesive granular systems~\cite{Gans2020,Pouliquen2025}. \note{As the liquid content increases, the liquid phase progressively forms higher-order clusters whose mechanical role remains poorly understood. The microscopic origin of the systematic increase in shear resistance with liquid content~\cite{Mitarai2006,Iveson2001} is not fully elucidated. In particular, it remains unclear how the geometry and spatial organization of liquid morphologies contribute to the capillary stress tensor, and how this microscopic stress translates into macroscopic rheological properties. Establishing such a morphology-resolved micro--macro connection is a key step toward predictive constitutive models for wet granular flows.}

\note{A central difficulty lies in the evolving morphology of the liquid phase. Under shear, liquid bridges break, reform, and coalesce into higher-order clusters~\cite{Mani2013}. Imaging studies have revealed spatial redistribution of liquid toward shear-zone and boundaries, as well as transitions between pendular and funicular morphologies~\cite{Awdi2025}. However, the quantitative contribution of these morphologies to the stress tensor has rarely been reconstructed experimentally. In most macroscopic models~\cite{Fisher1926,Rumpf1974}, cohesion is estimated from an average bridge force and coordination number, implicitly assuming isotropy and pendular connectivity. Such approaches capture the correct order of magnitude at small liquid content but cannot account for morphology transitions, anisotropic stress transmission, or shear-induced
redistribution.}

\note{On the theoretical side, a fully tensorial framework for capillary stress has been
formulated~\cite{Duriez2018}, expressing the capillary contribution to the Cauchy stress tensor in terms of interfacial geometry and capillary pressure. This framework predicts both isotropic and deviatoric components arising from the anisotropy of liquid interfaces. Yet direct experimental validation of this tensorial description --- resolved at the level of individual liquid domains under shear --- has remained limited.}

\note{Here, we bridge this gap by combining \emph{in situ} X-ray microtomography,
morphology-resolved segmentation, and micromechanical stress reconstruction in a sheared wet granular material. We reconstruct the full capillary stress tensor directly from the
three-dimensional geometry of each liquid morphology and quantify its contribution to the
macroscopic stress. This enables (i) a decomposition of the capillary pressure by morphology type, (ii) a direct comparison between microscopic capillary pressure and independently measured macroscopic friction, and (iii) an assessment of how shear-induced liquid redistribution modifies stress transmission.}

\note{Our results reveal that, despite the spatial accumulation of higher-order clusters near
shear-zone boundaries, the isotropic capillary pressure is overwhelmingly dominated by elementary two-grain bridges. Higher-order morphologies primarily affect spatial anisotropy rather than bulk cohesion. The microscopically reconstructed capillary pressure quantitatively predicts the increase of the macroscopic friction coefficient through an effective-stress closure without adjustable parameters. These findings provide experimental validation of tensorial capillary micromechanics under shear and establish a predictive micro--macro link for partially saturated granular flows \cite{Duriez2018}.}

% ============================================================
%  SECTION II — MATERIALS AND METHODS
% ============================================================
\section{Materials and Methods}
\label{sec:methods}

\subsection{Material and sample preparation}

Polystyrene beads of diameter $d = 580\,\mu\mathrm{m}$ (5\% polydispersity) and density
$\rho_s = 1050\,\mathrm{kg\,m^{-3}}$ are coated with a high-viscosity silicone oil
($\eta_f = 1093\,\mathrm{mPa\,s}$), with surface tension $\Gamma \simeq 20.6\,\mathrm{mN\,m^{-1}}$, density $\rho_L = 950\,\mathrm{kg\,m^{-3}}$, and nearly perfect wetting
($\theta \simeq 5^\circ$, $\cos\theta \simeq 0.996$)~\cite{Amarsid2024}. The liquid-to-solid volume ratio\footnote{This ratio $\epsilon$ is sometimes 
simply referred to as \emph{liquid content} in the present paper.} is set to 
$\epsilon = 0.03$, $0.05$, or $0.075$, spanning the pendular and early 
funicular regimes.

\subsection{Shear experiments and X-ray microtomography}

Shear experiments are performed using a custom-built device integrated into an X-ray
microtomography system, enabling \emph{in situ} imaging of the internal microstructure under controlled deformation~\cite{Awdi2025}. The granular material is confined in a transparent PMMA parallel-plate cup (diameter 20\,mm, gap 10\,mm). The bottom cup rotates while the upper plate remains fixed in rotation and imposes a constant normal stress by its weight ($\sigma_n \simeq 457\,\mathrm{Pa}$).

\note{After homogenization by two full rotations at a low shear rate ($\dot\gamma \simeq 0.17\,\mathrm{s^{-1}}$), the sample is sheared at a constant rate $\dot\gamma = 8.48\,\mathrm{s^{-1}}$, corresponding to an inertial number $I = \dot\gamma d/\sqrt{\sigma_n/\rho_s} \simeq 7\times10^{-3}$ and a viscous number $I_v = \eta_f\dot\gamma/\sigma_n \simeq 2\times10^{-2}$. Shear is interrupted after
prescribed intervals (2, 4, 6, 8, and 10\,min), and a full tomographic scan is acquired after each step.}

\note{Each scan is performed using a Varex 4343\,DX-I imager ($3032\times3032$ pixels,
139\,$\mu$m pixel pitch) and a Hamamatsu L10801 X-ray source (100\,kV, 80\,$\mu$A): 2368
projections are collected over a $360^\circ$ rotation, each averaged over three radiographs of 1\,s exposure, and reconstructed with RX Solutions' X-Act software into 3D volumes of $2517\times2519\times1260$ voxels at an 8\,$\mu$m voxel size. A full scan lasts approximately 2\,hours; owing to the high liquid viscosity, no measurable microstructural relaxation occurs during image acquisition.}

\note{The typical initial solid volume fraction is $\phi_0 \simeq 0.60$, consistent with a moderately dense packing. Under shear, a localized shear-zone develops near the upper no rotating boundary, accompanied by a slight local decrease of solid fraction due to dilatancy. The spatial extent of this shear-zone is determined independently from velocity profiles.}

\subsection{Segmentation and morphology classification}
Three-dimensional images are processed using an AI-based segmentation pipeline (Dragonfly
software~\cite{Dragonfly2022}), trained on semi-automatically labeled datasets. The model
reliably identifies grains, liquid domains, and background, after which liquid structures are automatically classified into capillary bridges (CB) and higher-order morphologies. Full details of the segmentation methodology and performance are provided in Ref.~\cite{Awdi2025} and summarized in Appendix~\ref{app:segmentation}.
\begin{figure}[h!]
\centering
\includegraphics[width=1\columnwidth]{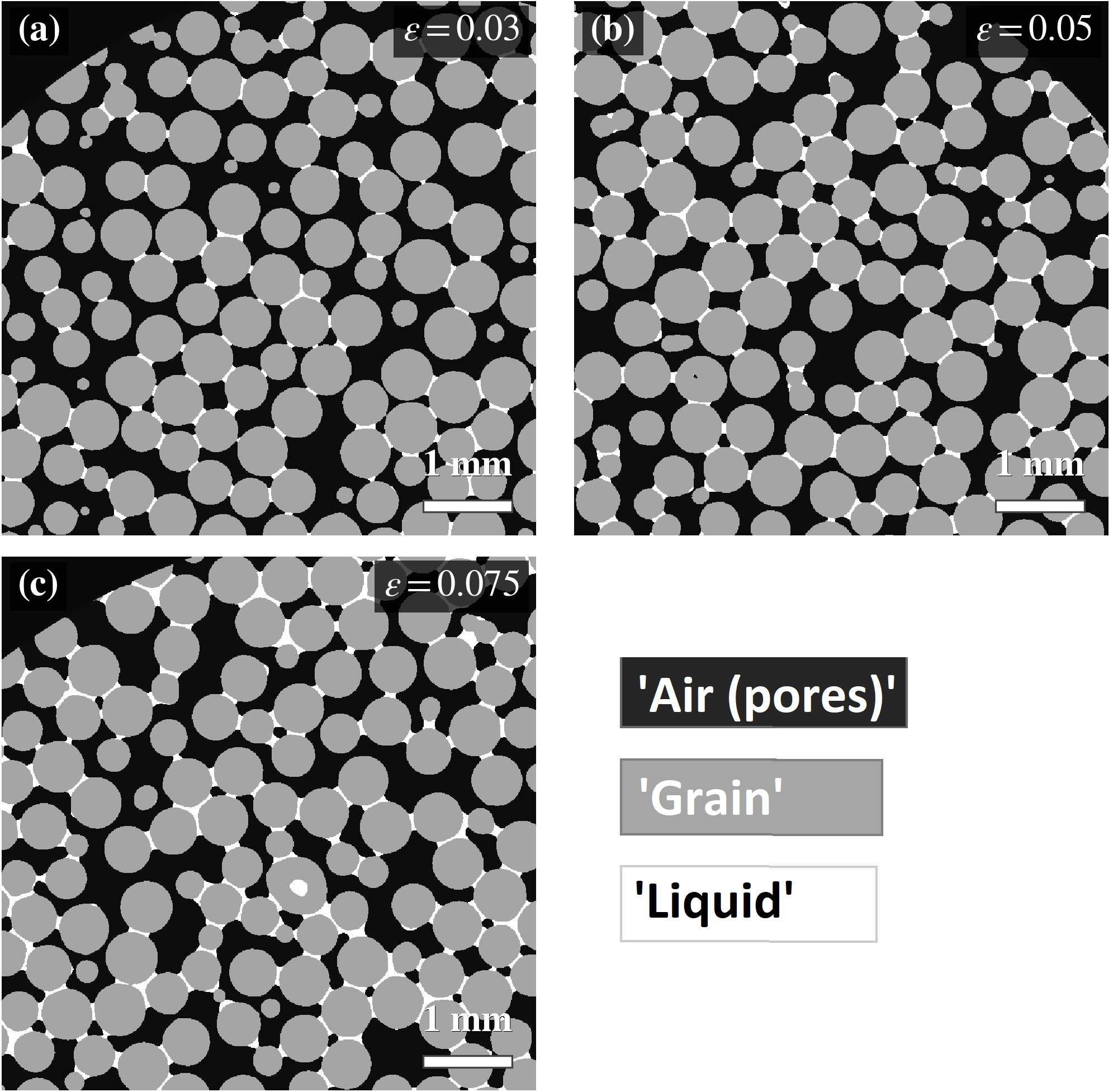}
\caption{2D cross-sections (6.70$\times$6.70~mm) from 3D segmented X-ray microtomographic images at zero shear deformation, for three values of \note{liquid content $\epsilon$:} (a)~$\epsilon=0.03$, (b)~$\epsilon=0.05$, and (c)~$\epsilon=0.075$.}
\label{fig:segmented images}
\end{figure}
Fig.~\ref{fig:segmented images} displays representative 2D slices extracted from the 3D segmented tomographic images. Each slice provides a cross-sectional view of the polystyrene beads and the liquid distribution at varying liquid contents ($\epsilon = 0.03, 0.05, \text{and } 0.075$). The images highlight the increasing prevalence of liquid clusters as the liquid content rises. At $\epsilon = 0.03$, discrete capillary bridges are clearly visible, connecting adjacent beads. As $\epsilon$ increases to $0.05$, the bridges become more interconnected, forming liquid agglomerations. Finally, at $\epsilon = 0.075$, the liquid phase tends to form more continuous, space-filling structures, enveloping multiple beads. Note that these images are taken at zero shear deformation, providing a baseline for comparison with images acquired under shear.

\begin{figure}[h!]
\centering
\includegraphics[width=.85\linewidth]{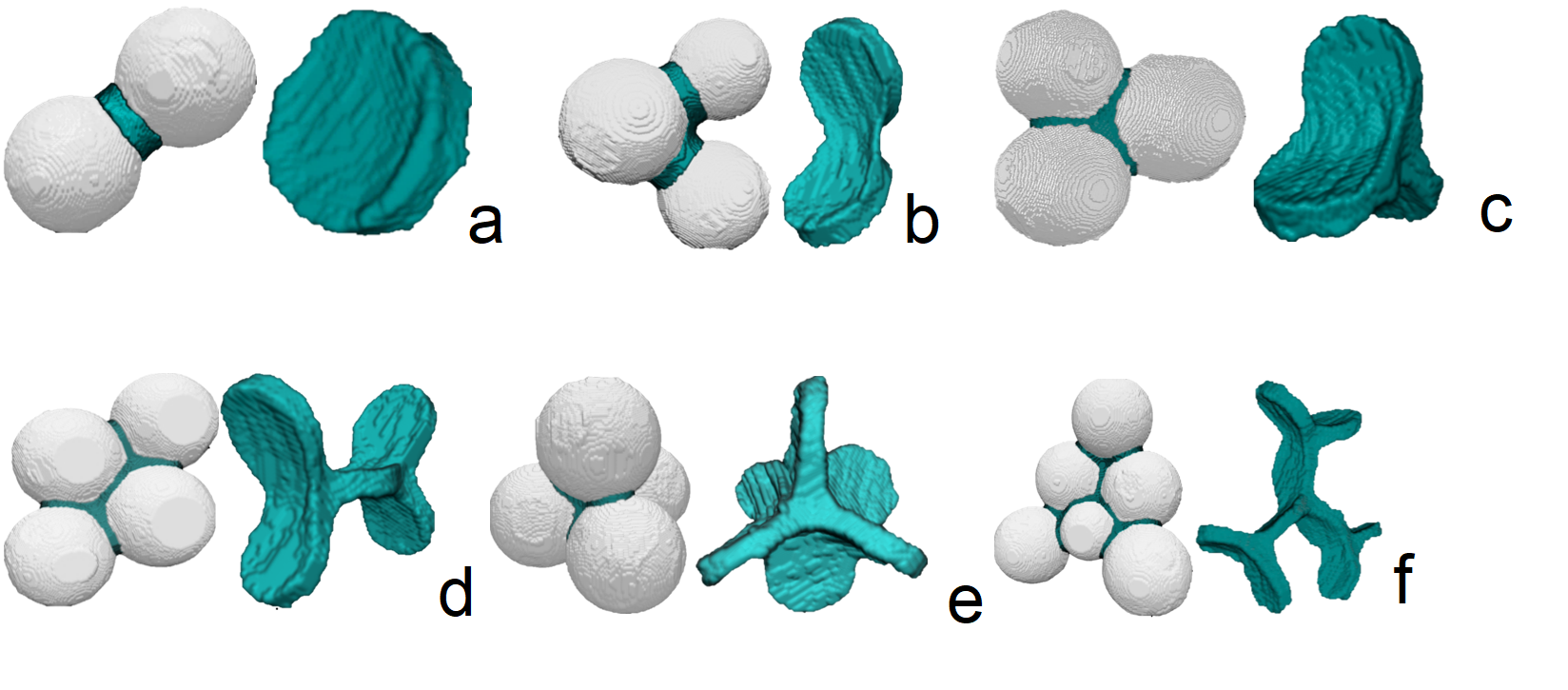}
\caption{Extracted liquid morphologies from the sample after segmentation, classified according to the number of coalesced CBs \note{and the number of connected grains: 
a) single CB; b) Dimer (2 and 3);  c) Trimer (3 and 3); d) Pentamer (5 and 4), e) Tetrahedral (6 and 4); f) Others: larger liquid clusters involving five or more  grains. From \cite{Awdi2025}}.}
\label{fig:3dmorphologies}
\end{figure}

\note{After segmentation, liquid domains are extracted as connected components and classified into isolated capillary bridges (CB, connecting a single grain pair), dimers (two connected CBs), trimers (three CBs), tetrahedral clusters (four CBs), pentamers (five CBs), and larger aggregates  Fig.~\ref{fig:3dmorphologies}. Validation against manually annotated three-dimensional sub-volumes demonstrates morphology-classification accuracy exceeding 95\% for liquid domains larger than 50 voxels. Such objects contribute negligibly to the total capillary stress and are excluded from the stress reconstruction.}

% ============================================================
%  SECTION III — MORPHOLOGY STATISTICS AND SPATIAL ORGANIZATION
% ============================================================
\section{Morphology Statistics and Spatial Organization}
\label{sec:morphology}

\subsection{Morphology distribution}

\begin{figure}[h]
    \centering
    \includegraphics[width=\linewidth]{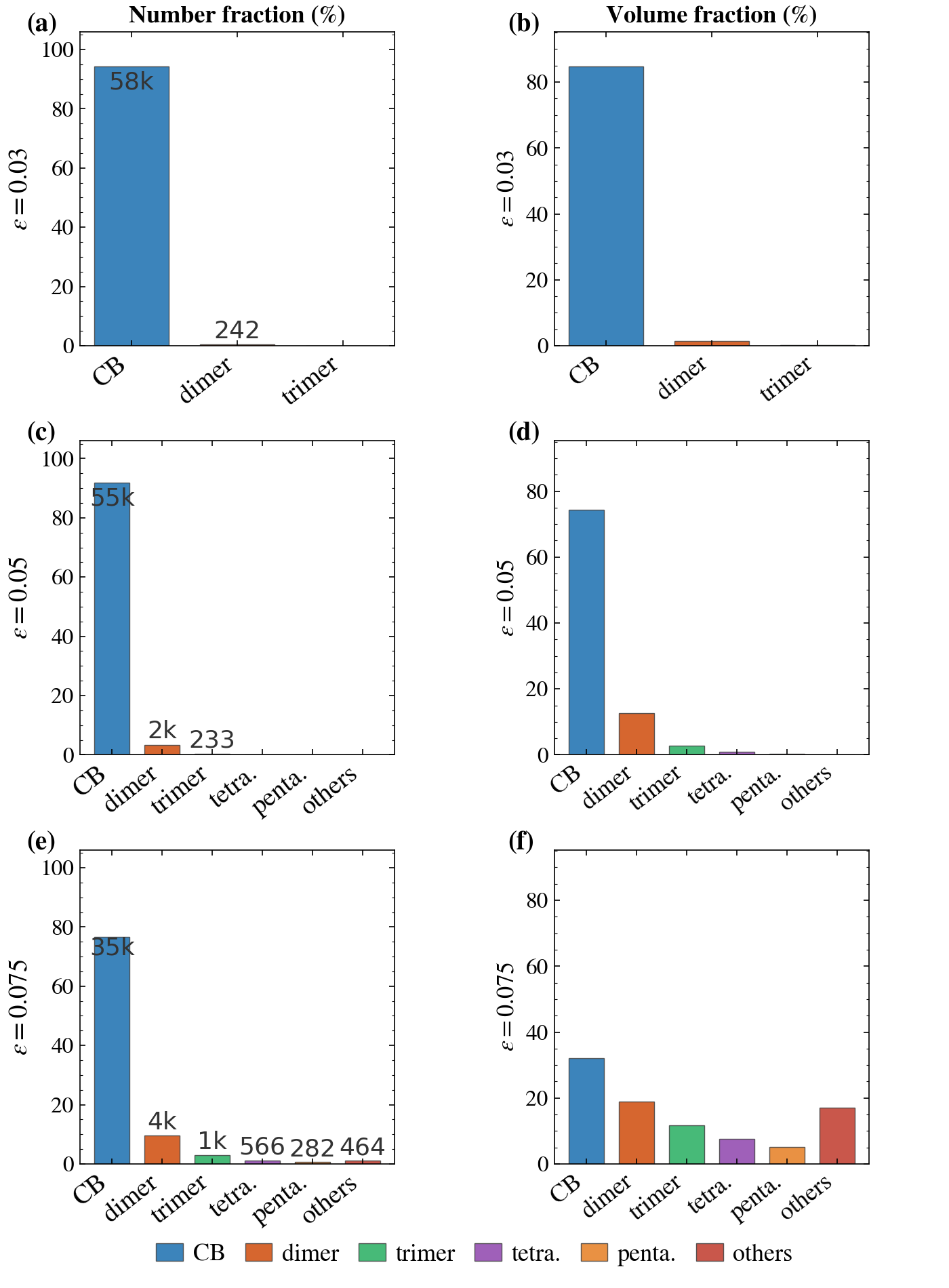}
    \caption{\note{Morphology-resolved statistics of the liquid domains for the three investigated liquid contents.
    Left column (a,c,e): number fraction of each morphology class (capillary bridge CB, dimer, trimer, tetrahedron, pentamer, and other clusters).
    Right column (b,d,f): fraction of the total liquid volume $V_w^\mathrm{tot}$ carried by each class.
    Rows correspond to $\epsilon=0.03$ (a,b), $0.05$ (c,d), and $0.075$ (e,f). Numbers above each bar indicate the absolute domain count.}}
    \label{fig:morphology_statistics}
\end{figure}

\note{Figure~\ref{fig:morphology_statistics} summarises the morphology distribution in terms of both domain count and liquid volume. At low liquid content ($\epsilon = 0.03$), capillary bridges (CB) overwhelmingly dominate: they account for 94\% of all liquid domains and carry 84\% of the total liquid volume. Dimers and trimers are scarce ($<1$\% by count).}

\note{As $\epsilon$ increases, the picture changes markedly in the volume channel. At $\epsilon = 0.05$, CB still represent 92\% of domains by count but their volume share decreases to 75\%, with dimers at 13\% and trimers at 3\%. The contrast is most striking at $\epsilon = 0.075$: CB retain 77\% by count, yet their volume fraction collapses to 32\%, with dimers (19\%), trimers (12\%), and larger clusters (17\%) each carrying a substantial volume fraction. This divergence between number and volume fractions signals the onset of the coalescence-dominated regime, in which a small number of large multi-grain clusters progressively replace the bridge-dominated microstructure as the primary liquid reservoirs.}

\subsection{Velocity profile and shear-zone geometry}

\begin{figure}[h!]
    \centering
    \includegraphics[width=1\linewidth]{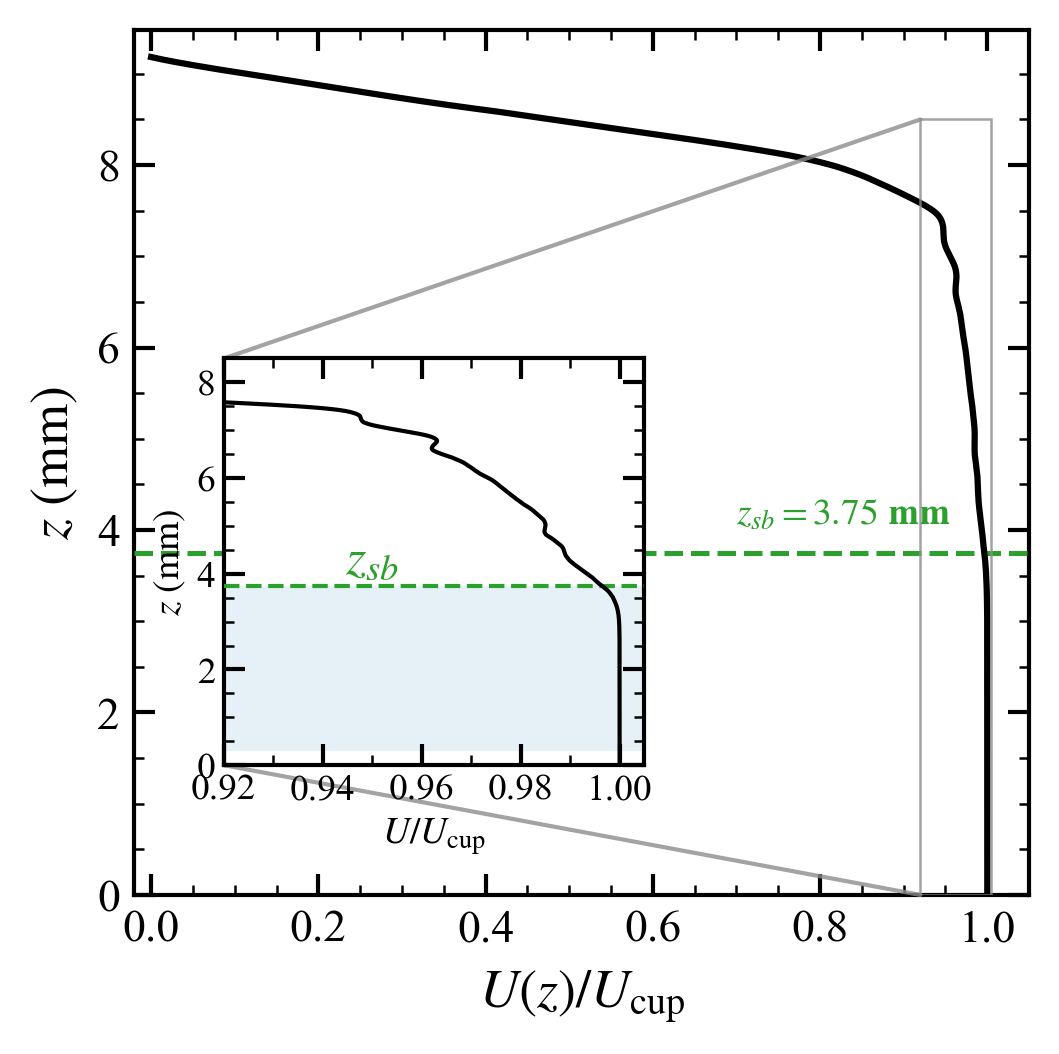}
    \caption{Vertical velocity profile at outer radius in sample with liquid content $\epsilon=0.03$ ($I=3.21\times10^{-3}$, $P^*=4.96$).  The velocity is normalized by the cup velocity $U_{cup}$ and the gap extends from the rotating cup at $z=0$ to the fixed upper plate at $z=z_{max}$. Inset: zoom on the low-velocity region, highlighting the dead zone (shaded) below the yielding threshold $z_{sb}\simeq3.75$~mm.}
    \label{fig:velocity}
\end{figure}

The vertical velocity profile $U(z)$ (Fig.~\ref{fig:velocity}) is extracted from side-view images via cross-correlation~\cite{Liu2015} (48\,px interrogation window, 50\% overlap, parabolic subpixel fitting). It exhibits a well-defined shear-zone near the rotating boundary, whose thickness ranges from $\delta \simeq 9d$ at low liquid fraction to $\delta \simeq 15d$ at the highest investigated saturation, in line with the reduced dilatancy and enhanced lubrication expected at higher saturation~\cite{Fiscina2012}. Outside this shear-zone, the material remains weakly sheared. \note{The shear-zone boundary $z_\mathrm{sb}$ is identified where $\partial U/\partial z$ drops below 10\% of its global maximum, yielding $z_\mathrm{sb} = 3.75 \pm 0.3$\,mm, with a shear-zone-thickness uncertainty of $\pm 0.5d$.}

\subsection{Spatial organization: density maps}

To quantify the spatial organization of liquid morphologies, we compute
absolute density maps $\rho_\mathrm{abs}(R,z)$, defined as the number of
liquid domains of class $m$ per unit volume of the annular bin,
\begin{equation}
  \rho_\mathrm{abs}(R,z) = \frac{N_\mathrm{bin}}{2\pi R\,\delta R\,\delta z},
  \label{eq:rho_abs}
\end{equation}
where $R$ is the radial bin center and $\delta R = \delta z = 0.96$\,mm
($= 120$~voxels).
\note{This annular normalisation yields a uniform density for a homogeneous
distribution at any $R$, correcting the geometric undercount that arises
from a rectangular-area normalisation at small $R$. Robustness tests (bin-size variations $\pm40$\%, removal of boundary slices) confirm that the localization patterns are insensitive to discretization choices (see Fig. \ref{fig:A2} in Appendix~\ref{app:density_maps}).}
\begin{figure}[h!]
  \centering
  \includegraphics[width=1\linewidth]{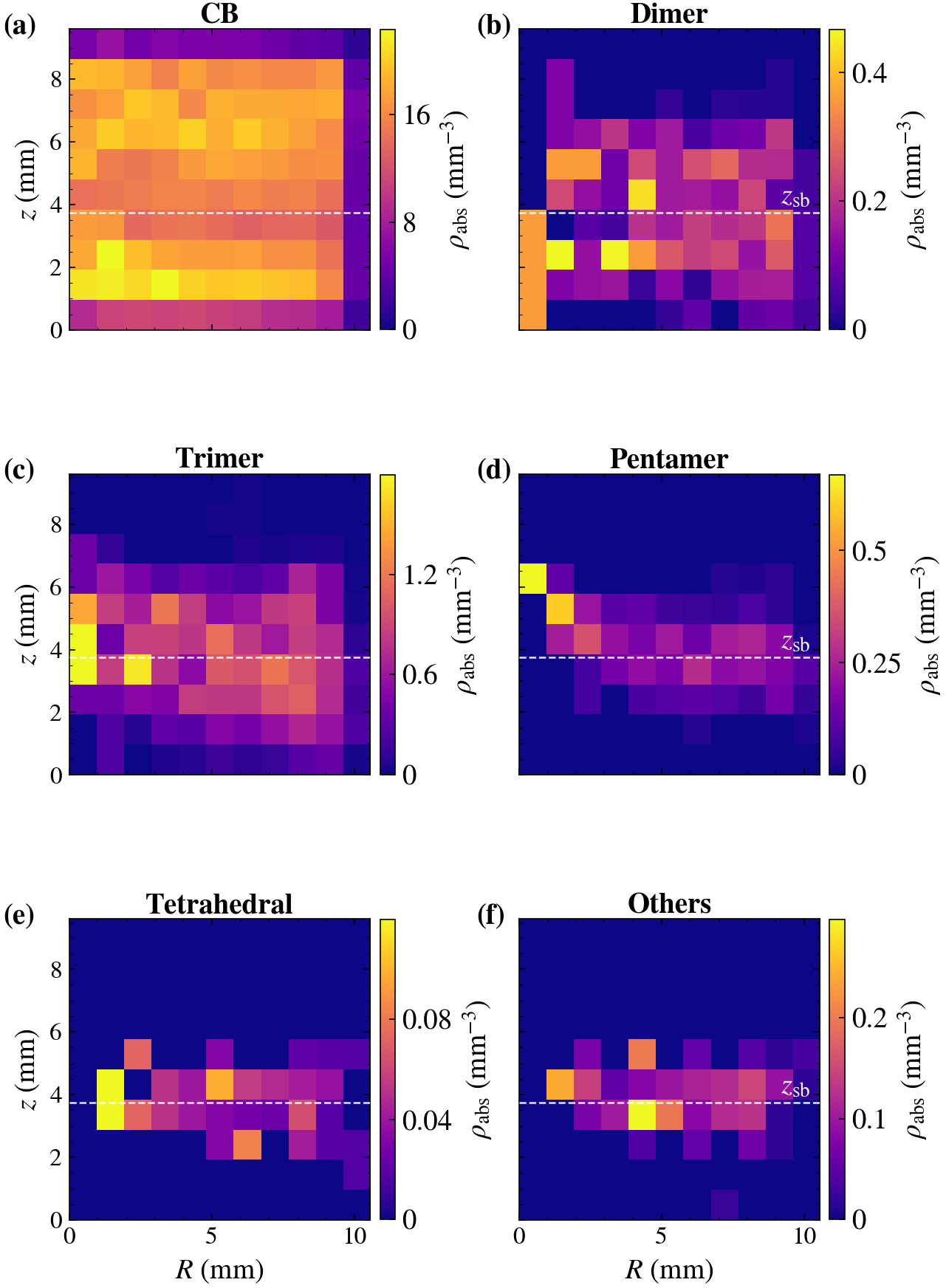}
  \caption{Absolute density maps $\rho_\mathrm{abs}(R,z)$ [mm$^{-3}$] of
  all liquid-domain morphology classes after 10\,min of shear
  ($\dot{\gamma}t = 5.098\times10^{3}$, $\epsilon = 0.05$).
  Densities are computed from centroid positions using
  Eq.~\eqref{eq:rho_abs} with $\delta R = \delta z = 0.96$\,mm
  ($= 120$~voxels); top and bottom slices ($|z| < 60$~voxels) are
  excluded to remove boundary artefacts.
  Each panel uses an individual colour scale (upper limit = 99th
  percentile of non-zero bin values); the white dashed line marks the
  shear-zone boundary $z_\mathrm{sb} \simeq 3.75$\,mm.}
  \label{fig:density_maps}
\end{figure}

Figure~\ref{fig:density_maps} shows the resulting density maps after 10\,min of shear ($\dot{\gamma}t = 5.098\times10^{3}$) for $\epsilon = 0.05$. Capillary bridges remain evenly distributed across the full sample height, confirming that the bridge population is not significantly altered by the shear flow. Dimers and trimers, by contrast, accumulate preferentially near the lower boundary of the shear zone ($z \lesssim z_\mathrm{sb} \simeq 3.75$\,mm), consistent with shear-driven liquid transport and local coalescence in the
high-strain region. Pentamers and larger morphologies are sparse and spatially confined, indicating that complex clusters form only under specific local flow
conditions near the shear-zone boundary.

\subsection{Shear-driven redistribution mechanism}

Shear localization governs the spatial redistribution of the liquid phase. Within the shear-zone, the large velocity gradients continuously fragment multi-grain clusters, leaving capillary bridges nearly uniform throughout the actively sheared region. In contrast, just below the shear-zone --- where strain rates are smaller --- liquid expelled from the shear-zone accumulates and promotes coalescence, producing the strong enrichment in dimers, trimers, and higher-order morphologies. The resulting asymmetric distribution of morphologies is therefore not random but directly governed by shear localization.

\begin{figure}[h!]
\centering
\includegraphics[width=1\linewidth]{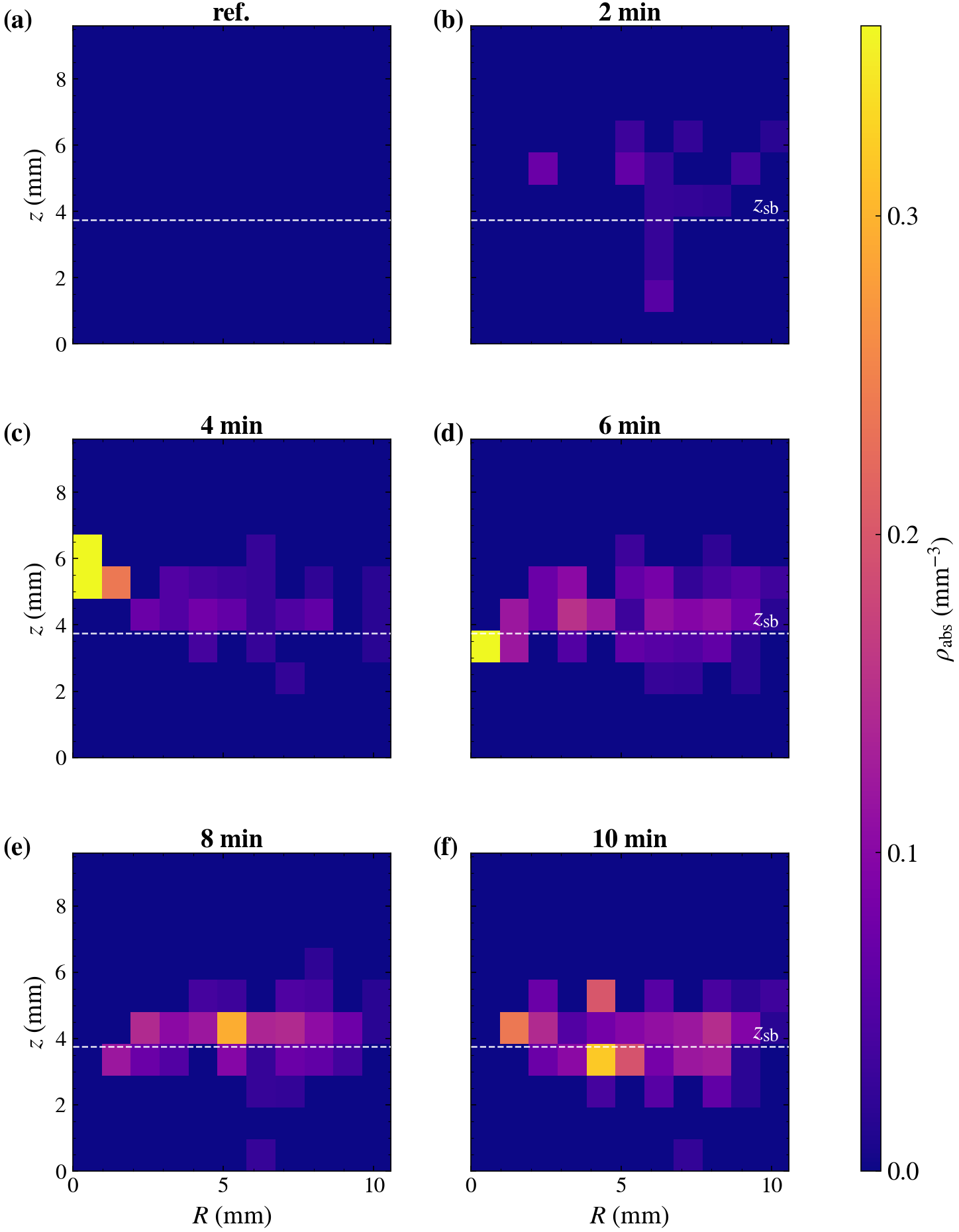}
\caption{Time evolution of the absolute density $\rho_\mathrm{abs}(R,z)$ [mm$^{-3}$] of the ``Others'' morphology class ($\varepsilon = 0.05$), from the reference state to
  10\,min of shear. All panels share a common colour scale (upper limit = 99th percentile of non-zero bin values); the white dashed line marks the shear-zone
  boundary $z_\mathrm{sb} \simeq 3.75$\,mm.}
\label{fig:others_temporal}
\end{figure}

The temporal evolution of the most complex clusters (``others'') confirms this mechanism
(Fig.~\ref{fig:others_temporal}). These structures are nearly absent in the unsheared reference state, appear transiently within the shear-zone at early deformation, and progressively accumulate outside the shear-zone as shear proceeds. Their growth correlates with liquid migrating out of the strongly sheared region, showing that higher-order morphologies result from shear-driven liquid transport rather than from local deformation within the shear-zone.

At the highest liquid fraction ($\epsilon = 0.075$, see Fig. \ref{fig:A5} \& \ref{fig:A6} in Appendix~\ref{app:coalescence}),
tomography reveals a qualitative change in the shear-induced liquid dynamics. Instead of
fragmenting large clusters, shear expels liquid from the active shear-zone and promotes
coalescence in the adjacent weakly sheared region, resulting in a systematic depletion of small morphologies within the shear-zone and the progressive emergence of large, connected liquid clusters outside it. This competition marks the onset of a transition from a bridge-dominated (pendular) regime toward a coalescence-dominated one, where liquid connectivity increases and collective liquid structures play an increasing geometrical role.

\subsection{Coarse-grained structural maps}

\note{Local solid and liquid volume fractions are obtained by convolving the segmented 3D binary volumes with a uniform cubic kernel of side $N_k = 120$ voxels ($\approx 1.0\,\mathrm{mm}\approx 1.7d$), projected onto the $(R,z)$ plane by radial averaging in annular bins of width $\Delta R = 0.2$\,mm. The outermost 20 voxels at top and bottom are excluded to avoid convolution boundary artefacts, and the analysis is restricted to
$0.5 \leq R \leq 9.0$\,mm. The local liquid-to-solid ratio $\varepsilon(\mathbf{x}) = \phi_l/\phi_s$ is defined only where $\phi_s > 0.10$.}
\begin{figure}[h!]
\centering
\includegraphics[width=1\linewidth]{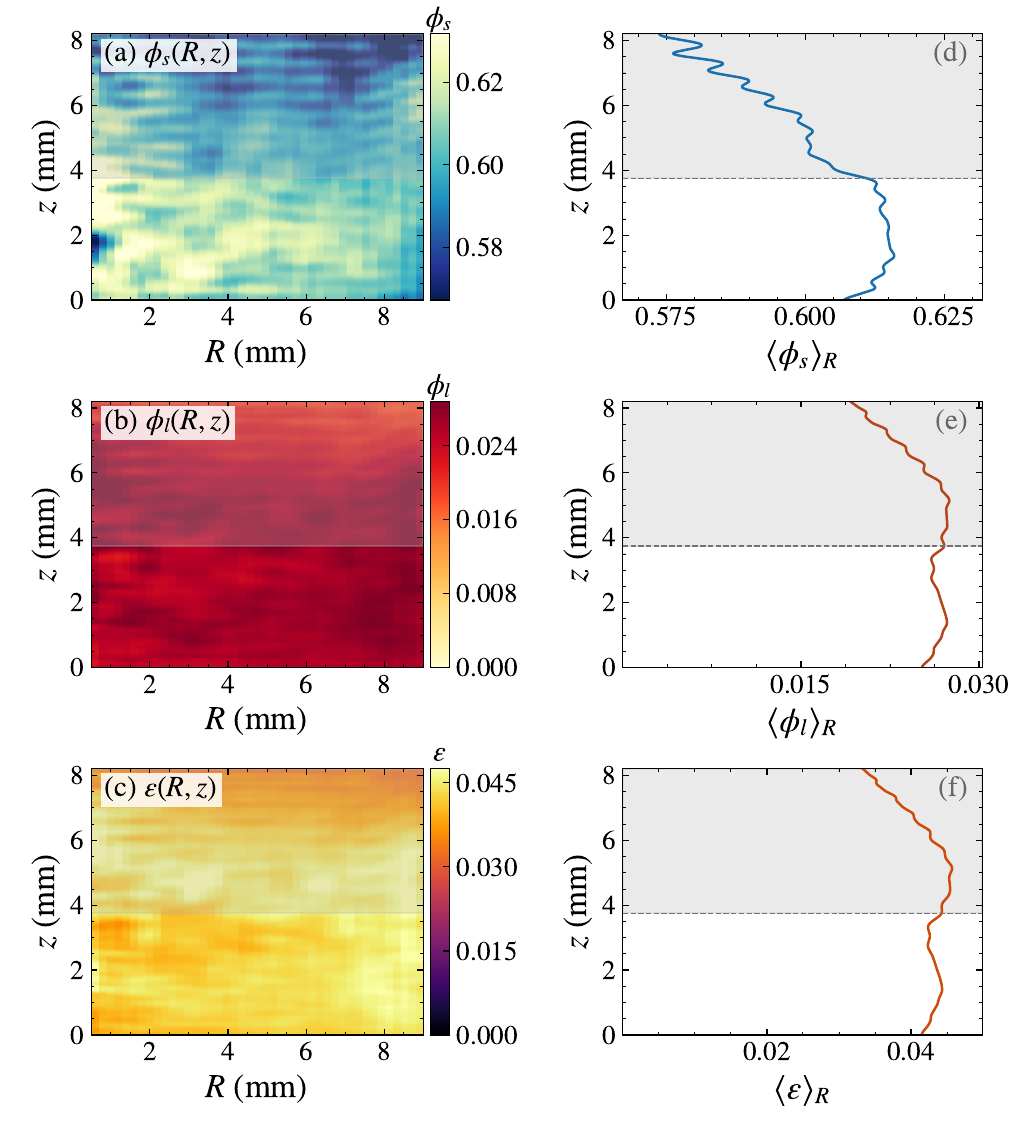}
\caption{Coarse-grained structural maps for $\varepsilon = 0.05$ after $8$~min of shear, obtained by convolving the segmented 3D volume with a cubic kernel of side $N_k = 120$~voxels (${\approx}\,1.0~\mathrm{mm} \approx 1.7\,d$). Left column: 2D maps as functions of radial position $R$ and axial position $z$. Right column: corresponding radially averaged profiles $\langle\,\cdot\,\rangle_R(z)$. The shaded region ($z \geq 3.75~\mathrm{mm}$, dashed line) marks the shear-zone (see Fig.~\ref{fig:velocity}). (a,\,d)~Solid volume fraction $\phi_s$. (b,\,e)~Liquid volume fraction $\phi_l$. (c,\,f)~Local liquid-to-solid ratio $\varepsilon = \phi_l/\phi_s$. }
\label{fig:structural_maps}
\end{figure}

\note{As shown in Fig.~\ref{fig:structural_maps}, the solid fraction $\phi_s(R,z)$ decreases in the shear-zone ($z \gtrsim 3.75$\,mm), from $\langle\phi_s\rangle_R \approx 0.62$ in the bulk to $\approx 0.57$--$0.58$ ($\Delta\phi_s/\phi_s \approx -8\%$), consistent with shear-induced dilatancy. Simultaneously, the liquid volume fraction $\phi_l(R,z)$ also decreases in the shear-zone ($\Delta\phi_l/\phi_l \approx -35$--$40\%$), indicating that liquid is expelled from the shear zone toward the unsheared region. As a result, $\varepsilon = \phi_l/\phi_s$ decreases from $\langle\varepsilon\rangle_R \approx 0.047$--$0.050$ in the bulk to
$\approx 0.030$--$0.032$ in the shear-zone, dominated by liquid expulsion rather than by
dilatancy ($|\Delta\phi_l/\phi_l| \approx 35$--$40\%$ vs.\ $|\Delta\phi_s/\phi_s|\approx 8\%$).}

\note{The redistribution is primarily axial: within the bulk ($2 \lesssim R \lesssim 7~\mathrm{mm}$), both $\phi_s$ and $\varepsilon$ are approximately uniform in $R$ at a given height $z$, consistent with the axisymmetric geometry of the plate-cup cell. Residual radial gradients near the axis ($R \lesssim 1~\mathrm{mm}$) and the outer wall ($R \gtrsim 8~\mathrm{mm}$) are associated with confinement effects; these boundary regions are excluded from the morphology analysis.}

\note{The morphology maps further reveal three key features: (i) capillary bridges are ubiquitous and remain structurally stable under shear; (ii) multi-grain liquid clusters emerge through shear-driven transport and local coalescence; (iii) the spatial organization of the liquid phase is controlled by shear localization rather than by homogeneous saturation.}

% ============================================================
%  SECTION IV — CAPILLARY STRESS RECONSTRUCTION
% ============================================================
\section{Capillary Stress Reconstruction}
\label{sec:stress}

\subsection{Micromechanical framework}

The total Cauchy stress tensor may be written as
\begin{equation}
\bm{\sigma}_\mathrm{total} = \bm{\sigma}_\mathrm{contact} + \bm{\sigma}_\mathrm{capillary},
\label{eq:cauchy}
\end{equation}
where $\bm{\sigma}_\mathrm{contact}$ arises from grain--grain contact forces and
$\bm{\sigma}_\mathrm{cap}$ from capillary interactions transmitted through liquid domains.

Following the micromechanical framework of Duriez \emph{et al.}~\cite{Duriez2018}, the capillary
stress associated with a given liquid morphology $k$ occupying a representative volume $\Omega$ is
\begin{equation}
\bm{\sigma}^{(k)}_\mathrm{cap} = -\frac{1}{\Omega}
\Bigl[u_c\bigl(\bm{\mu}^{(k)}_{Vw} + \bm{\mu}^{(k)}_{Ssw}\bigr)
+ \Gamma\bigl(\bm{\mu}^{(k)}_{Snw} + \bm{\mu}^{(k)}_{\Gamma}\bigr)\Bigr],
\label{eq:sigma_cap_full}
\end{equation}
where $u_c$ is the capillary pressure inside the liquid domain, $\Gamma$ the gas--liquid surface
tension, and $\Omega$ the sample volume. The four microstructural tensors (defined in
Appendix~\ref{app:tensors}) encode the geometry of the fluid domain: liquid volume for
$\bm{\mu}_{Vw}$, solid--wetting interface for $\bm{\mu}_{Ssw}$, gas--liquid interface for
$\bm{\mu}_{Snw}$, and triple line for $\bm{\mu}_{\Gamma}$.

\note{Because silicone oil nearly perfectly wets the polystyrene beads, the contributions of $\bm{\mu}_{Ssw}$ and $\bm{\mu}_\Gamma$ are negligible (see Appendix~\ref{app:tensors}).}
Equation~\eqref{eq:sigma_cap_full} thus reduces to
\begin{equation}
\bm{\sigma}^{(k)}_\mathrm{cap} = -\frac{1}{\Omega}
\Bigl[u_c\,\bm{\mu}^{(k)}_{Vw} + \Gamma\,\bm{\mu}^{(k)}_{Snw}\Bigr].
\label{eq:sigma_cap_reduced}
\end{equation}

\subsection{Numerical implementation}

\note{For each liquid domain, the gas--liquid interface is reconstructed via the Marching Cubes algorithm on a Gaussian-smoothed binary mask ($\sigma = 1.5$ voxels), and the interface tensor $\bm{\mu}_{Snw}$ is evaluated by numerical integration over the resulting triangulated surface~\cite{Schott2025}:}
\begin{equation}
\mu^{(k)}_{Snw,ij} = \int_{S^{(k)}_{nw}} (\delta_{ij} - n_i n_j)\,\mathrm{d}S.
\label{eq:mu_snw}
\end{equation}
\note{The capillary pressure $u_c$ is obtained independently from the measured domain volume using the toroidal pendular bridge model (Appendix~\ref{app:toroidal}), which is resolution-independent and does not require resolving the meniscus curvature directly.}

\note{The morphology-resolved capillary stress tensor is thereby assembled without invoking any pairwise-force model --- an important distinction from standard DEM approaches, which apply pairwise capillary forces between grain pairs and are therefore limited to isolated pendular bridges.}

\subsection{Morphology-resolved capillary pressure budget}

\note{The total capillary stress is obtained by summing over all liquid domains:
\begin{equation}
\bm{\sigma}_\mathrm{capillary} = \sum_k \bm{\sigma}^{(k)}_\mathrm{cap}.
\label{eq:sigma_total}
\end{equation}}
The isotropic component of each morphology-resolved tensor defines a capillary pressure
$P^{(k)}_\mathrm{cap} = \tfrac{1}{3}\mathrm{Tr}(\bm{\sigma}^{(k)}_\mathrm{cap}) < 0$.

Summing over all liquid domains yields a microscopic capillary pressure
$P^\mathrm{micro}_\mathrm{cap} \simeq -52.3$\,Pa for $\epsilon = 0.05$, normalized by the full sample volume $\Omega_\mathrm{tot} = \pi R^2 H$. Nearly 83\% of this pressure originates from simple two-grain bridges, while dimers, trimers, and larger clusters contribute collectively only $\sim 17\%$, despite their preferential accumulation near the shear-zone boundary.

% ============================================================
%  SECTION V — EFFECTIVE-STRESS CLOSURE AND COMPARISON WITH RHEOLOGY
% ============================================================
\section{Effective-Stress Closure}
\label{sec:closure}

\note{To connect the microscopic capillary pressure to the macroscopic rheology, we invoke the effective-stress approach validated numerically and experimentally by Badetti \emph{et
al.}~\cite{Badetti2018b}. In that work, DEM simulations of polystyrene beads wetted by silicone oil showed that the quasistatic yield condition of the wet assembly is well described by that of the dry material subjected to an effective normal stress augmented by the isotropic capillary pressure. The resulting closure for the apparent friction coefficient reads
\begin{equation}
\mu^*_w = \mu_\mathrm{dry}\!\left(1 + \frac{|P^\mathrm{micro}_\mathrm{cap}|}{\sigma_n}\right),
\label{eq:closure}
\end{equation}
where $\mu_\mathrm{dry} = 0.257$ is the friction coefficient of the dry material in the quasistatic limit~\cite{Badetti2018b}, and $\sigma_n \simeq 457$\,Pa is the applied normal stress. This relation is equivalent to the Rumpf--Coulomb cohesion formula in the quasistatic limit~\cite{Rumpf1974,Badetti2018b}: the capillary pressure $P^\mathrm{micro}_\mathrm{cap}$ plays the role of an isotropic effective
confining stress that augments the frictional resistance of the dry assembly.}

\begin{figure}[h!]
    \centering
    \includegraphics[width=0.40\textwidth]{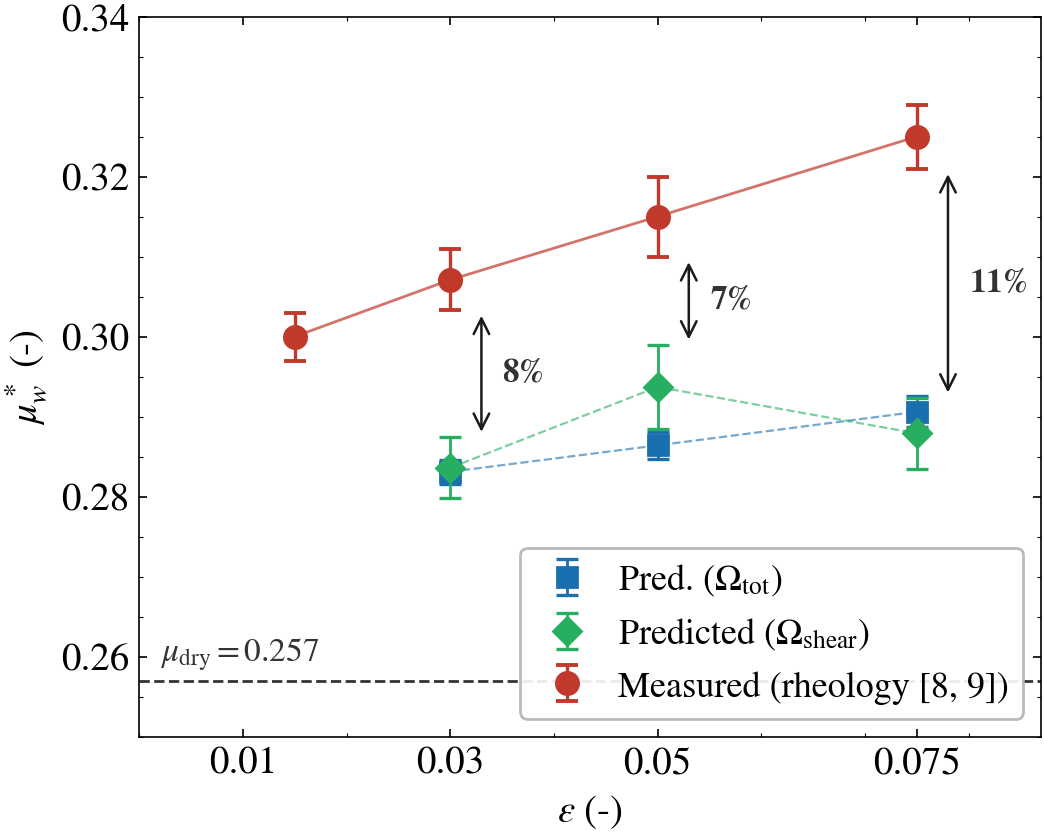}
    \caption{Effective-stress closure for wet granular shear flow. Symbols: measured friction coefficient $\mu_w^*$ (red circles, rheometry~\cite{Badetti2018a, Badetti2018b}); predicted values from the
    closure Eq.~\eqref{eq:closure} using $\Omega_\mathrm{tot}$ (blue squares) and $\Omega_\mathrm{shear}$ (green diamonds). Error bars on predictions reflect combined uncertainties in $u_c$ ($\pm5\%$, Lian model), liquid-domain volume ($\pm3\%$, voxelization), and, for $\Omega_\mathrm{shear}$, the shear-zone height ($\pm0.5$~mm out of $3.75$~mm). The dry friction coefficient $\mu_\mathrm{dry}=0.257$ is shown as a dashed line. Both predictions underestimate the measured $\mu_w^*$ by $5$--$11\%$, with $\Omega_\mathrm{shear}$ providing the closer agreement; the true effective cohesive volume lies between the two limits.}
    \label{fig:closure}
\end{figure}

\note{The capillary pressure entering Eq.~\eqref{eq:closure} should reflect the cohesion in the region that controls the rheological response, namely the shear-zone. When $P^\mathrm{micro}_\mathrm{cap}$ is evaluated with volume $\Omega_\mathrm{tot}$, the predicted friction coefficient $\mu^*_w \approx 0.283$--$0.291$ systematically underestimates the measured values $0.307$--$0.325$ by 8--11\% (Fig.~\ref{fig:closure}, blue squares). Restricting the sum to liquid domains located within the shear-region ($\Omega_\mathrm{shear} = \pi R^2 z_\mathrm{shear}$, $z_\mathrm{shear} = 3.75$\,mm) instead yields $\mu^*_w \approx 0.284$--$0.294$, reducing the underestimation to 5--7\% (Fig.~\ref{fig:closure}, green diamonds). The residual gap suggests that the effective cohesive volume lies between $\Omega_\mathrm{shear}$ and $\Omega_\mathrm{tot}$,
corresponding to an effective height $z_\mathrm{eff} \approx 4$--$5$\,mm consistent with the extent of the dilatancy zone.}

\note{A note of caution is warranted regarding the comparison with the Badetti closure~\cite{Badetti2018b}. That closure was calibrated on DEM simulations in which all liquid bridges are geometrically identical, homogeneously distributed, and follow a fixed volume fraction --- conditions that correspond to an idealized pendular microstructure with no higher-order morphologies. In the present experiment, however, the actual liquid-domain distribution is morphologically heterogeneous: dimers, trimers, and larger clusters coexist with capillary bridges and are spatially localised near the shear-zone boundary (Section~\ref{sec:morphology}).
The closure relation of Eq.~\eqref{eq:closure} therefore captures only the scalar (isotropic) part of the capillary stress, and cannot account for the deviatoric contributions that arise from the anisotropic spatial organisation of higher-order morphologies.}

\note{The Badetti closure captures the qualitative trend over the full range of investigated liquid contents: the systematic increase of $\mu^*_w$ with $\epsilon$ is reproduced by both predictions, confirming that the frictional enhancement is driven by the capillary pressure budget.}

\note{The residual gap points to two compounding effects. First, averaging $P^\mathrm{micro}_\mathrm{cap}$ over $\Omega_\mathrm{tot}$ dilutes the local cohesive contribution acting on the shearing plane, since liquid bridges outside the shear-zone do not directly resist the applied shear stress --- an effect corrected, at least partially, by restricting the averaging volume to $\Omega_\mathrm{shear}$. Second, at the highest liquid content ($\epsilon = 0.075$), the coalescence-dominated regime produces higher-order clusters whose anisotropic spatial distribution introduces a deviatoric
capillary stress $\sigma^\mathrm{cap}_{12}$ that contributes positively to the apparent friction but is absent from the isotropic effective-stress closure of Eq.~\eqref{eq:closure}. Both effects act in the same direction, explaining why $\Omega_\mathrm{shear}$ provides consistently closer agreement and why the underestimation is largest at $\epsilon = 0.075$.}

\note{Nevertheless, the discrepancy remains moderate ($\leq 11\%$), confirming that pendular bridges continue to dominate the cohesive budget and that the effective-stress closure provides a reliable first-order prediction across the full range of investigated saturation levels.}

% ============================================================
%  SECTION VI — CONCLUSION
% ============================================================
\section{Conclusion}
\label{sec:conclusion}

\note{We have combined \emph{in situ} X-ray microtomography, morphology-resolved segmentation, and micromechanical stress reconstruction to establish a quantitative micro--macro description of shear rheology in wet granular materials across the pendular and early funicular regimes.}

\note{At the morphological level, shear localization is the primary driver of liquid redistribution. Capillary bridges are ubiquitous and structurally stable under shear, while higher-order clusters --- dimers, trimers, and larger aggregates --- emerge through shear-driven liquid transport and preferentially accumulate near the lower boundary of the shear-zone. At the highest liquid fraction ($\epsilon = 0.075$), this mechanism transitions toward a coalescence-dominated regime in which large connected clusters progressively replace the bridge-dominated microstructure outside the active shear-zone.}

\note{At the stress level, the morphology-resolved reconstruction reveals that, despite the spatial prominence of higher-order clusters near the shear-zone boundary, the isotropic capillary pressure is overwhelmingly controlled by elementary two-grain bridges, which account for nearly 85\% of $P^\mathrm{micro}_\mathrm{cap}$ at $\epsilon = 0.05$. Higher-order morphologies primarily affect the spatial anisotropy of the liquid phase rather than the bulk cohesion. Inserting the reconstructed capillary pressure into the effective-stress closure of Badetti \emph{et al.}~\cite{Badetti2018b} quantitatively reproduces the trend of the macroscopic friction coefficient $\mu^*_w$ across the full range of liquid contents, with a residual underestimation of 5--11\% attributed to the finite extent of the effective cohesive volume and, at high saturation, to deviatoric capillary contributions from anisotropic higher-order clusters.}

\note{These findings provide experimental validation of tensorial capillary micromechanics under shear, and establish that %the Rumpf--Coulomb effective-stress picture remains predictive 
the effective-stress closure of Eq.~\eqref{eq:closure} — equivalent to
the Rumpf--Coulomb cohesion formula in the quasistatic limit — remains
predictive even when the liquid phase partially transitions beyond the strict pendular regime. A natural extension of this work would be to investigate higher saturation levels, where cluster percolation and funicular connectivity are expected to modify both the stress transmission mechanism and the spatial organization of cohesion --- effects that lie beyond the reach of pairwise-additive capillary models and that may require continuum descriptions of the interstitial liquid phase.}

\begin{acknowledgments}
Funding from the Agence Nationale de la Recherche (ANR) is gratefully acknowledged under the
Investments for the Future program ANR-11-LABX-022-01 and the project RheoGranoSat
(ANR-16-CE08-0005-01). A.F.\ thanks O.\ Pitois, A.\ Lemaitre, B.\ Dollet, and F.\ Schott for
insightful discussions.
\end{acknowledgments}

% ============================================================
%  DATA AVAILABILITY
% ============================================================
\section*{Data and code availability}

Representative segmented volumes, morphology catalogues, and Python analysis scripts are
available upon reasonable request to the corresponding author (abdoulaye.fall@cnrs.fr).
Surface-integration routines for computing $\bm{\mu}_{Snw}$ and $\bm{\sigma}_\mathrm{cap}$ are
available upon reasonable request. %A public repository (Zenodo) is in preparation and will be referenced in the published version.

% ============================================================
%  APPENDICES
% ============================================================
\appendix

% -----------------------------------------------------------
\section{AI-assisted segmentation and morphology classification}
\label{app:segmentation}
% -----------------------------------------------------------

The 3D tomograms were processed using an AI-assisted segmentation pipeline specifically developed --- to identify grain, liquid, and air voxels --- and validated for unsaturated wet granular materials~\cite{Awdi2025}. The workflow combines a Random-Forest classifier for voxel-level pre-segmentation to generate ground-truth slices on which a U-Net convolutional network is trained, reaching a 97--98\% segmentation accuracy estimated by the Dice coefficient.
Resolution-related uncertainties preclude a precise measurement of the liquid volume, but they are confined to a few voxels near the air--liquid or solid--liquid interfaces and do not affect the identification of liquid morphologies at the grain scale.

\begin{figure}[h]
  \centering
  \includegraphics[width=0.49\linewidth]{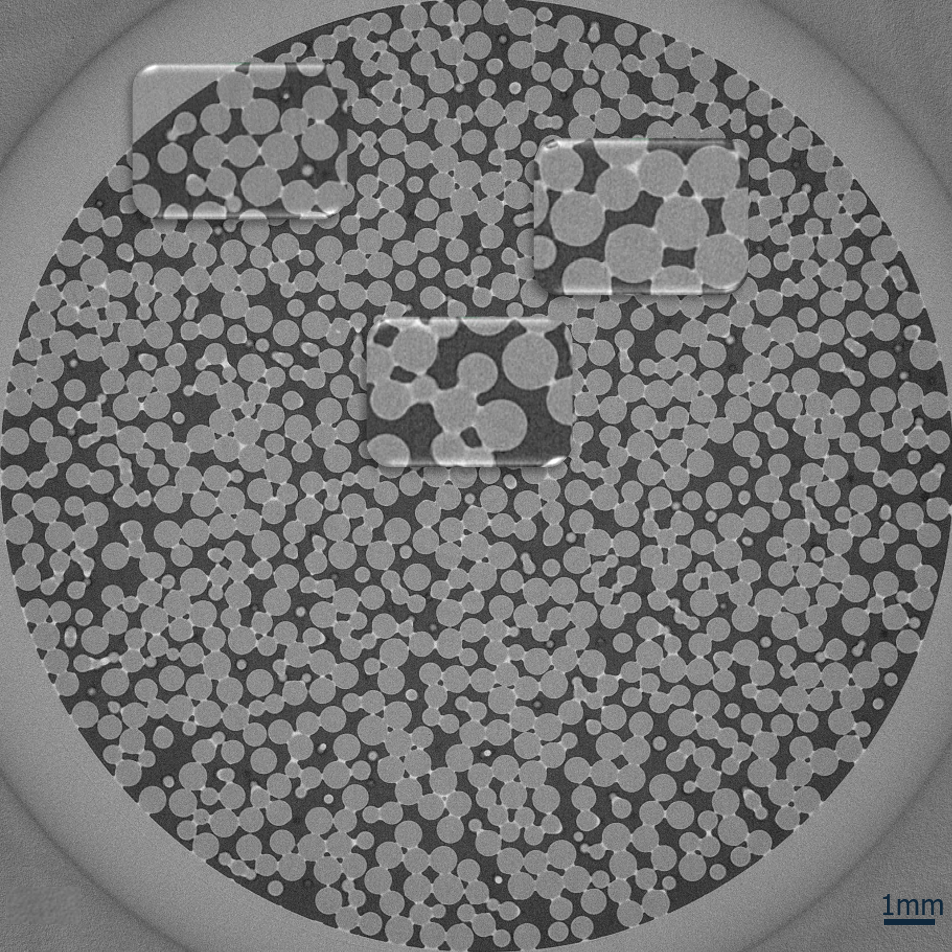}
\includegraphics[width=0.49\linewidth]{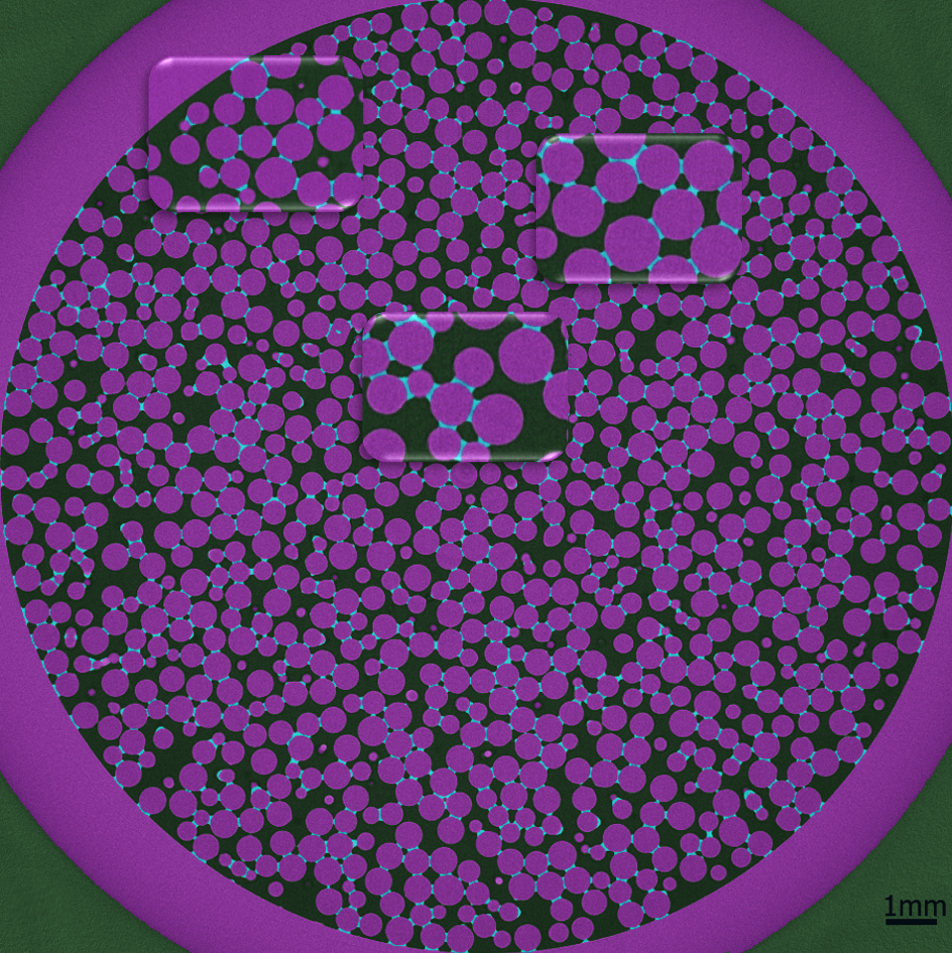}
  \caption{ (Left) Raw X-ray tomographic slice (8\,$\mu$m resolution) before segmentation. (Right) Representative segmented horizontal slice with grain voxels shown in violet, air (pore space) in black, and liquid in cyan. From~\cite{Awdi2025}.} 
  \label{fig:A1}
\end{figure}

After segmentation, liquid domains are extracted as connected components and classified into capillary bridges (CB), dimers, trimers, tetrahedral clusters, pentamers, and larger aggregates through an automated procedure. The method identifies the grains belonging to each liquid body and assigns a morphology label based on the number and geometry of grain--grain connections identified by dilation of the liquid domain into neighboring watershed grain regions.

To assess the reliability of the morphology classification, a dedicated validation was performed against manually annotated three-dimensional sub-volumes extracted from representative tomographic datasets. These sub-volumes were selected to span different shear times and spatial locations, including both the shear-zone and the weakly sheared bulk. For liquid domains with volumes exceeding 50 voxels, the classification accuracy exceeds 95\% for all morphology classes. Below this threshold, liquid objects approach the voxel resolution and their shape is dominated by discretization noise, making reliable morphological identification intrinsically ambiguous. These small objects are excluded from the validation analysis and from the capillary-stress computation; their contribution to the total capillary stress is negligible ($<0.5\%$ of $P^\mathrm{micro}_\mathrm{cap}$).

% -----------------------------------------------------------
\section{Robustness of absolute-density maps}
\label{app:density_maps}
% -----------------------------------------------------------

\begin{figure}[h!]
  \centering
  \includegraphics[width=1\linewidth]{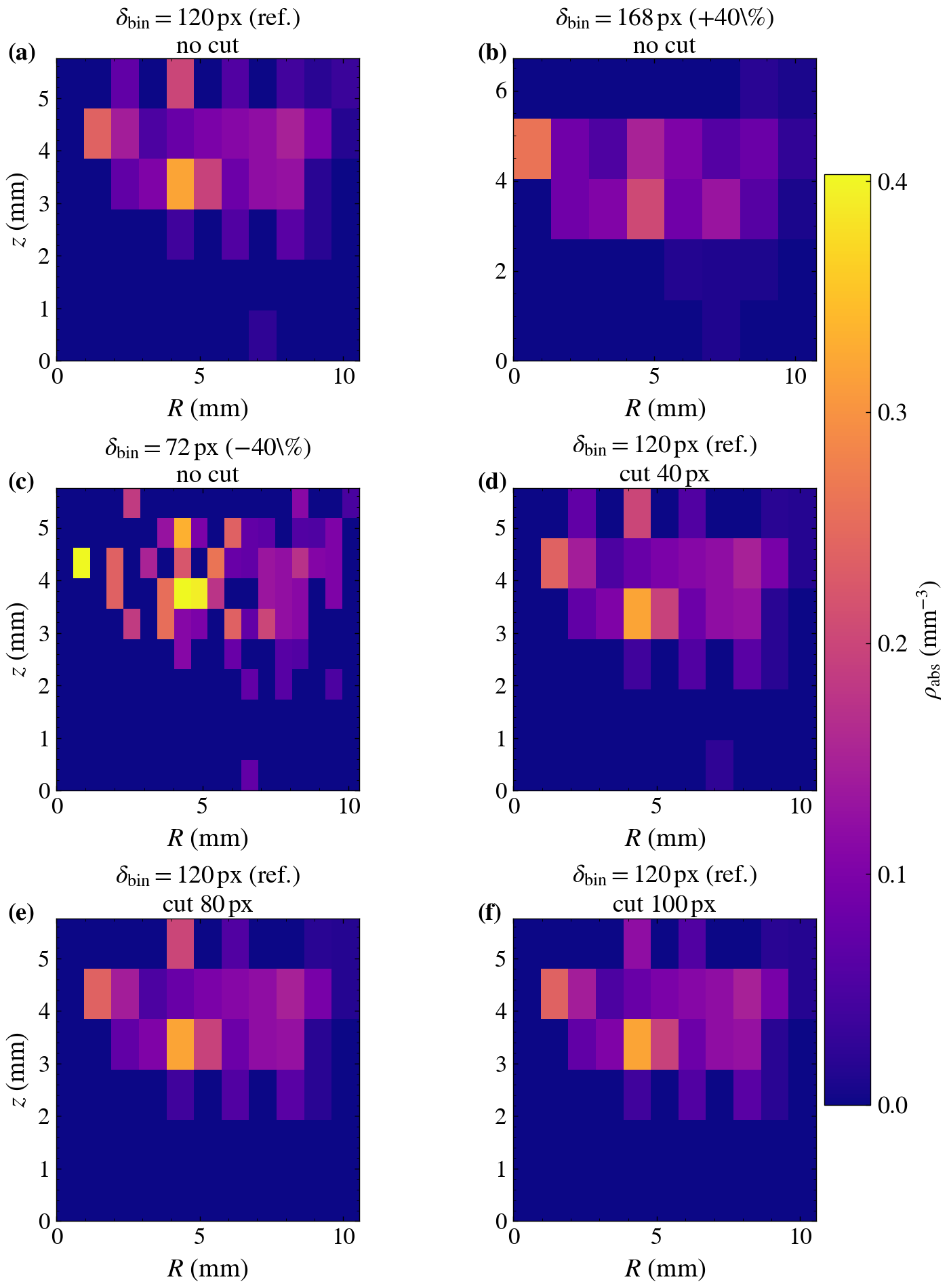}
  \caption{ Absolute density maps $\rho_\mathrm{abs}(r,z)$ of the ``Others'' morphology at $t = 10$\,min ($\epsilon = 0.05$), computed using different analysis parameters. Top row: reference bin size (120\,px) and variations of $\pm 40\%$ with no axial cut. Bottom row: reference bin size with increasing removal of top and bottom slices (40, 80, and 100\,px) to  exclude boundary artefacts. Peak amplitudes vary by less than 6\% and the spatial localization pattern is unchanged in all cases.}
  \label{fig:A2}
\end{figure}

The absolute-density maps $\rho_\mathrm{abs}(r,z)$ reported in the main text are constructed from the centroid positions of all liquid morphologies using uniform radial--axial binning
($\Delta r = \Delta z = 120\,\mathrm{px} = 0.96\,\mathrm{mm}$). To verify that the localization patterns observed for dimers, trimers, and higher-order morphologies are not numerical artifacts, we performed a series of robustness checks (Fig. \ref{fig:A2}):
\begin{itemize}
\item Varying the bin size by $\pm40\%$ modifies peak amplitudes by less than 6\%, without changing the position or shape of localized regions.
\item Removing 40--80 slices from the top and bottom of the stack (to exclude boundary effects from the filter kernel) leaves the spatial organization inside the shear-zone unchanged.
\item 1D axial and radial projections of $\rho_\mathrm{abs}(r,z)$ reproduce the same localization trends across all shear times.
\end{itemize}

\begin{figure}[h]
  \centering
  \includegraphics[width=1\linewidth]{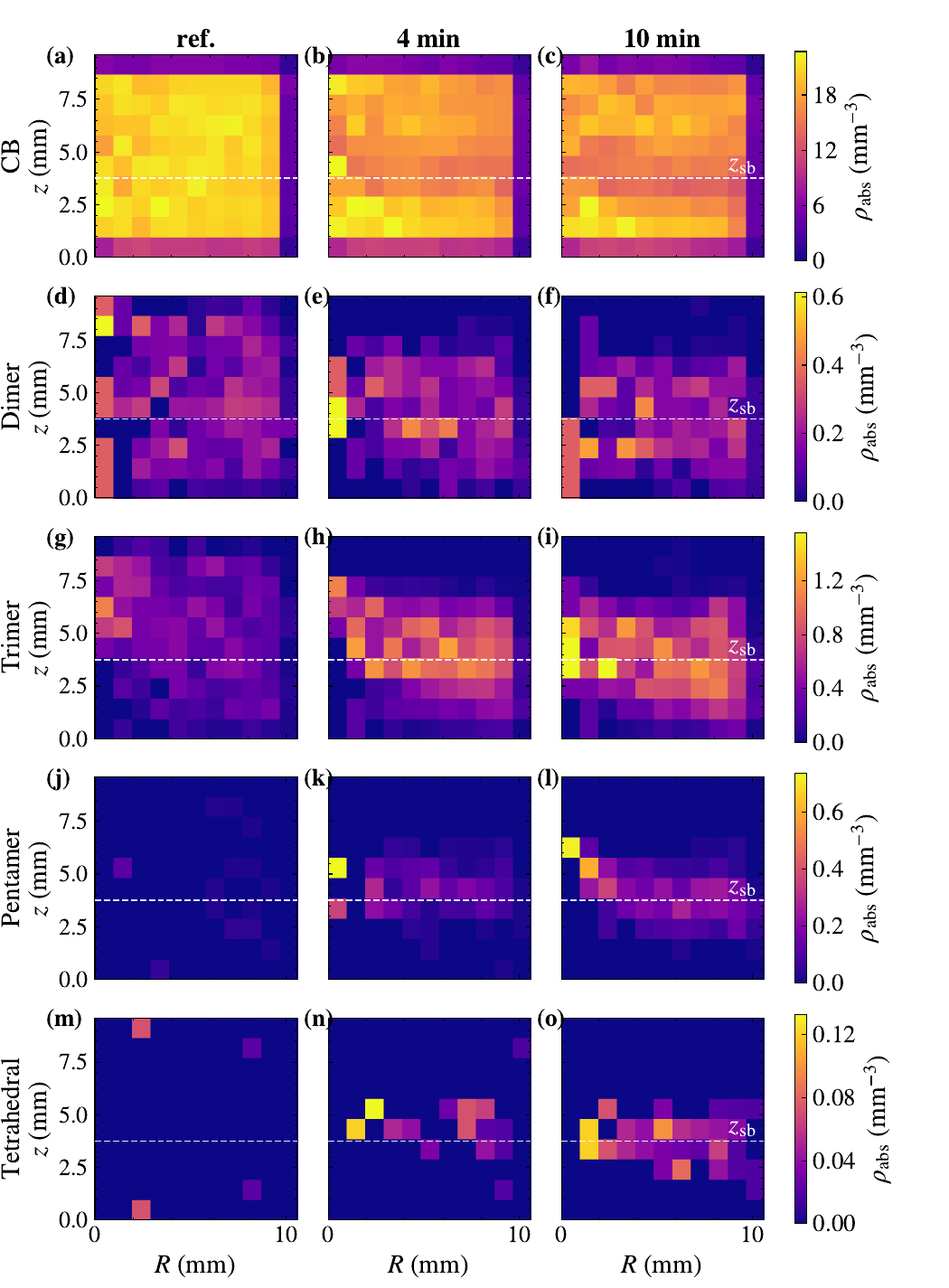}
    \caption{Time-resolved absolute-density maps $\rho_\mathrm{abs}(R,z)$ of liquid-domain morphologies at $\epsilon = 0.05$, from the reference state (before shear) to 10\,min of shear. Rows correspond to capillary bridges~(CB), dimers, trimers, pentamers, and tetrahedral clusters; columns show three
  representative times (reference, 4\,min, 10\,min). Each row shares a common colour scale (upper limit = 99th percentile of non-zero values, to avoid saturation by isolated outliers).
  %
  %Each row shares a common colour scale (per-morphology maximum,
  %$p_{99}$); 
  The white dashed line marks the shear-zone boundary at $z_\mathrm{sb}\simeq 3.75$\,mm. Bin size $\delta_\mathrm{bin} = 120$\,px ($= 0.96$\,mm);
  boundary slices $|z| < 60$\,px removed on each side to
  exclude segmentation artefacts.}
  \label{fig:A3}
\end{figure}

Representative time-resolved maps for capillary bridges, dimers, and trimers (at $\epsilon = 0.05$) confirm that capillary bridges remain uniformly distributed throughout the sample at all times, while dimers and trimers progressively accumulate near the lower boundary of the shear-zone ($z \simeq 3.75$\,mm) (Fig.\ref{fig:A3}). Despite their lower absolute density, pentamers and tetrahedral clusters exhibit spatial organization equally robust with respect to bin-size variations and boundary-slice removal.

Hence, the observed growth and spatial localization of multi-grain morphologies are robust with respect to numerical discretization and reflect genuine shear-induced liquid redistribution. 

% -----------------------------------------------------------
\section{High liquid fraction: shear-induced coalescence}
\label{app:coalescence}
% -----------------------------------------------------------

At the highest liquid fraction ($\epsilon = 0.075$), tomography reveals a qualitative change in
the effect of shear on liquid morphology. Rather than fragmenting large clusters, shear expels
liquid from the active shear-zone and promotes coalescence in the adjacent weakly sheared region.
This leads to a progressive depletion of small morphologies inside the shear-zone and the
emergence of large, connected liquid clusters outside it. 
\begin{figure}[h]
  \centering
  \includegraphics[width=1\linewidth]{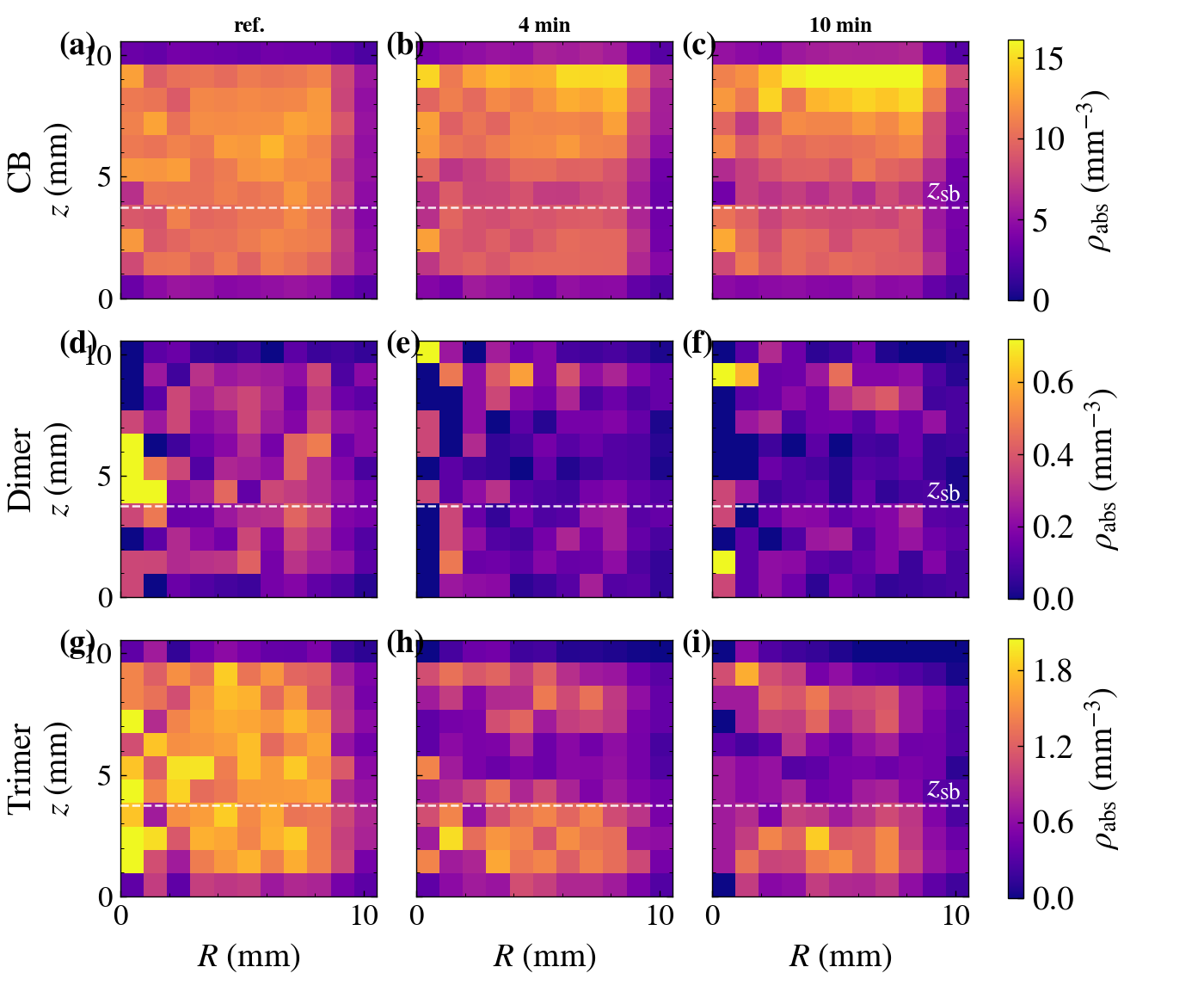}
  \caption{Time-resolved absolute-density maps
  $\rho_\mathrm{abs}(R,z)$ of capillary bridges~(CB), dimers, and
  trimers at $\epsilon = 0.075$, for the reference state,
  4\,min, and 10\,min of shear.
  Each row shares a common colour scale (per-morphology $p_{99}$);
  the white dashed line marks the shear-zone boundary
  $z_\mathrm{sb}\simeq 3.75$\,mm.
  From $t\geq 4$\,min, all three morphologies show a progressive
  depletion inside the shear zone ($z\gtrsim z_\mathrm{sb}$),
  consistent with shear-driven liquid expulsion from the active
  flow region.
  Outside the shear zone ($z\lesssim z_\mathrm{sb}$), the density
  of dimers and trimers increases with time, indicating that the
  expelled liquid reorganises into larger connected structures
  in the weakly sheared region.}
  \label{fig:A5}
\end{figure}

\begin{figure}[h]
  \centering
  \includegraphics[width=\linewidth]{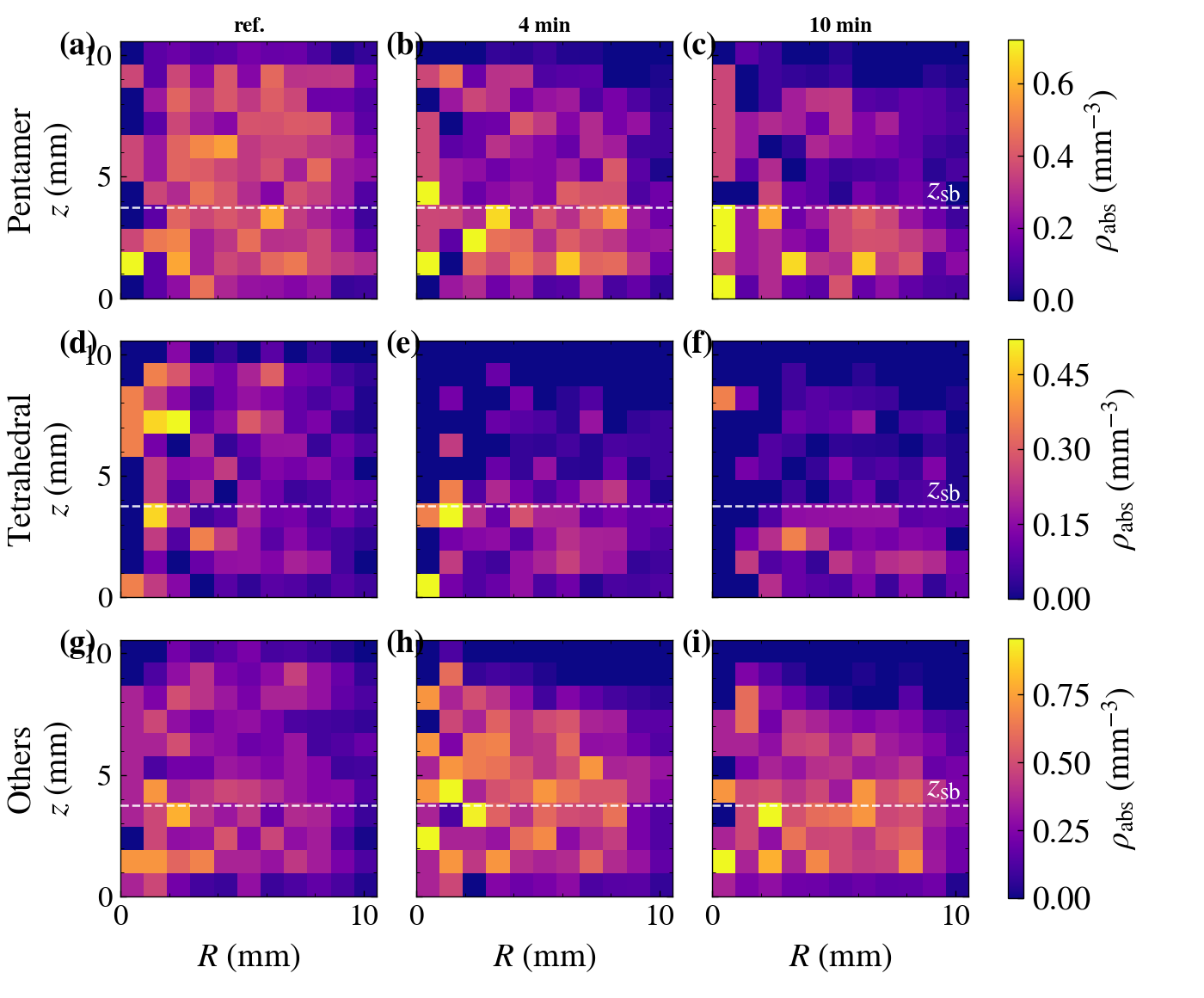}
  \caption{Same as Fig.~\ref{fig:A5} for pentamers, tetrahedral
  clusters, and the ``Others'' class (clusters of six or more grains)
  at $\epsilon = 0.075$.
  From $t \geq 6$\,min, the density of large clusters grows markedly
  outside the shear zone ($z \lesssim z_\mathrm{sb}$), while remaining
  low or absent inside it ($z \gtrsim z_\mathrm{sb}$).
  This spatial segregation --- large clusters accumulating in the weakly
  sheared bulk while the active shear zone is depleted of liquid ---
  marks the onset of the coalescence-dominated regime and is the
  structural counterpart of the larger discrepancy observed in the
  effective-stress closure at $\epsilon = 0.075$
  (see Section~\ref{sec:closure}).}
  \label{fig:A6}
\end{figure}

The progressive depletion of CB, dimer, and trimer morphologies (Fig. \ref{fig:A5}) in the shear zone ($z \lesssim 3.75$\,mm) is visible from $t \geq 4$\,min, while the growth of large clusters (pentamers, tetrahedrals, ``Others'') outside the shear-zone from $t \geq 6$\,min marks the onset of the coalescence-dominated regime (Fig. \ref{fig:A6}). This behavior is discussed in the main text in connection with the associated limitations of the effective-stress closure at $\epsilon = 0.075$.

% -----------------------------------------------------------
\section{Physical meaning of the four microstructural tensors}
\label{app:tensors}
% -----------------------------------------------------------

Following the micromechanical formulation of Duriez \emph{et al.}~\cite{Duriez2018}, the morphology-resolved capillary stress associated with a liquid domain reads:
\begin{equation}
\bm{\sigma}_\mathrm{cap} = -\frac{1}{\Omega}
\Bigl[u_c\bigl(\bm{\mu}_{Vw} + \bm{\mu}_{Ssw}\bigr)
+ \Gamma\bigl(\bm{\mu}_{Snw} + \bm{\mu}_{\Gamma}\bigr)\Bigr],
\label{eq:app_sigma_full}
\end{equation}
where $u_c$ is the capillary pressure, $\Gamma$ the gas--liquid interfacial tension, and $\Omega$ the sample volume. As detailed below, two of the four tensors ($\bm{\mu}_{Ssw}$ and $\bm{\mu}_\Gamma$) are negligible for the present system ($\theta \simeq 5^\circ$), and the stress reconstruction reduces to:
\begin{equation}
\bm{\sigma}_\mathrm{cap} \approx -\frac{1}{\Omega}
\Bigl[u_c\,V_w\,\mathbf{I} + \Gamma\,\bm{\mu}_{Snw}\Bigr],
\label{eq:app_sigma_reduced}
\end{equation}
where the volume tensor $\bm{\mu}_{Vw}$ has been replaced by its isotropic part $V_w\mathbf{I}$ ($V_w$ being the liquid domain volume), since the deviatoric correction from the second moment of the liquid volume is small compared to the $\bm{\mu}_{Snw}$ contribution for the morphologies investigated here.

\paragraph{Wetting-fluid volume tensor.}
\begin{equation}
\mu_{Vw,ij} = \int_{V_w} x_i x_j\,\mathrm{d}V.
\label{eq:mu_Vw}
\end{equation}
This second-moment tensor describes the spatial distribution of the wetting fluid. Its isotropic part is $V_w\mathbf{I}/3$; its deviatoric part reflects the geometric elongation of the liquid domain. In practice, for compact liquid morphologies (CB, dimers, trimers), the deviatoric contribution of $\bm{\mu}_{Vw}$ is smaller than 5\% of the total stress and is absorbed into the $\bm{\mu}_{Snw}$ term in the reduced form.

\paragraph{Wetted solid--fluid surface tensor.}
\begin{equation}
\mu_{Ssw,ij} = \int_{S_{sw}} n_i n_j\,\mathrm{d}S.
\label{eq:mu_Ssw}
\end{equation}
This tensor accounts for the orientation of liquid-wetted solid surfaces. For nearly perfectly wetting systems ($\theta \lesssim 5^\circ$), $\cos\theta \approx 1$ and this term enters Eq.~\eqref{eq:app_sigma_full} with a prefactor $u_c(1-\cos\theta) < 0.004\,u_c$, making its contribution to the total stress negligible ($<1\%$). It is accordingly set to zero in the stress reconstruction.

\paragraph{Gas--liquid interface tensor.}
\begin{equation}
\mu_{Snw,ij} = \int_{S_{nw}} (\delta_{ij} - n_i n_j)\,\mathrm{d}S.
\label{eq:mu_Snw}
\end{equation}
This tensor captures the anisotropy of the gas--liquid interface through the complement of the dyadic product of the outward normal $\mathbf{n}$. It is the dominant source of deviatoric capillary stress, directly linking the meniscus geometry to shear resistance. It is evaluated numerically by integrating over the triangulated gas--liquid surface obtained from the Marching Cubes reconstruction after Gaussian pre-smoothing ($\sigma = 1.5$ voxels) to remove staircase artifacts from the voxelized interface.

\paragraph{Triple-line tensor.}
\begin{equation}
\mu_{\Gamma,ij} = \int_{\Gamma} \ell_i\ell_j\,\mathrm{d}\ell,
\label{eq:mu_Gamma}
\end{equation}
where $\bm{\ell}$ is the unit tangent to the three-phase contact line. This tensor vanishes identically for $\theta = 0$. For the present system ($\theta \simeq 5^\circ$, $\cos\theta \simeq 0.996$), its prefactor $\Gamma\sin\theta \approx 0.087\,\Gamma$ renders its contribution negligible; it is not included in the stress reconstruction.

Together, these four tensors provide a complete geometric description of the liquid domain that determines how its morphology generates capillary stresses transmitted to the granular skeleton, without invoking any pairwise-force approximation.

% -----------------------------------------------------------
\section{Capillary pressure: toroidal bridge model}
\label{app:toroidal}
% -----------------------------------------------------------

The capillary pressure $u_c$ in Eq.~\eqref{eq:sigma_cap_reduced} is estimated from the measured liquid-domain volume $V_w$ using the toroidal pendular bridge model of Lian \emph{et al.}~\cite{Lian1993}. For contacting grains ($h=0$, i.e.\ dimensionless separation $S^*=0$), the toroidal meniscus geometry is fully characterized by the half-filling angle $\phi$ and the contact angle $\theta$.
  \begin{figure}[h!]
  \centering
  \includegraphics[width=.8\linewidth]{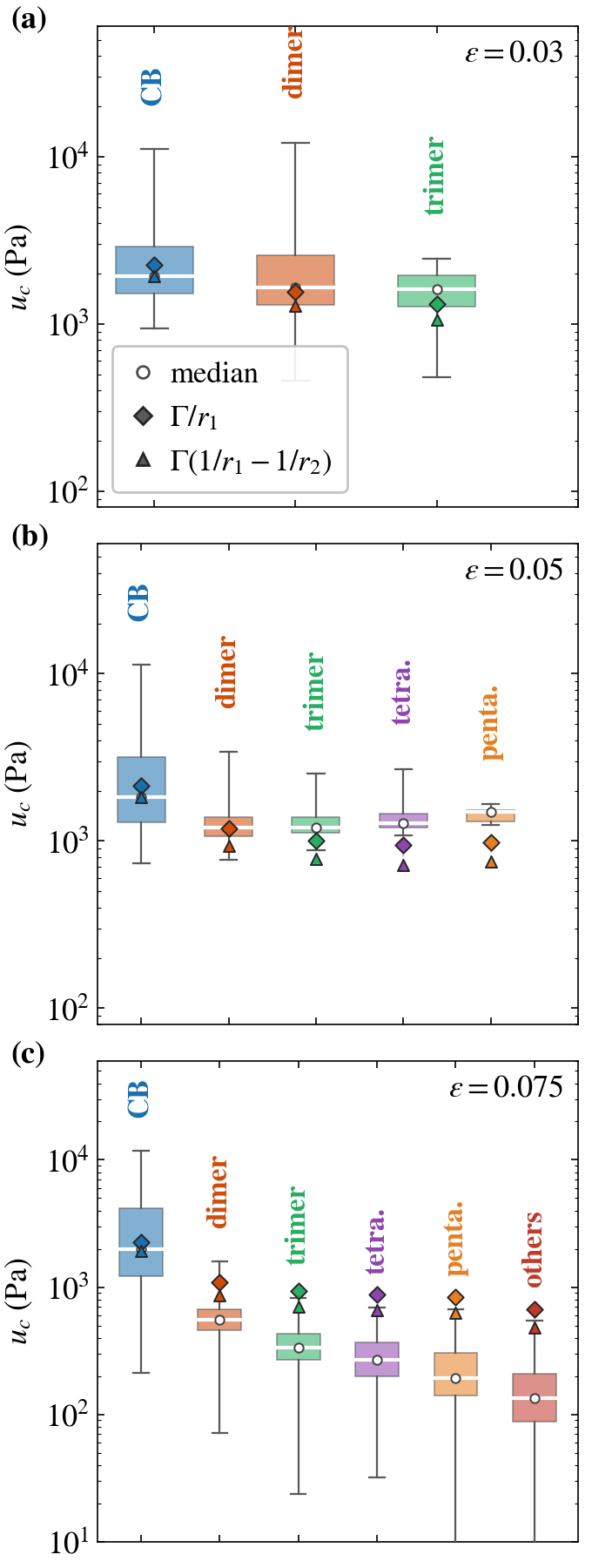}
  \caption{Distribution of the capillary pressure $u_c$ computed from the toroidal bridge model~\cite{Lian1993} for each liquid-domain morphology class and liquid content $\epsilon$. Box: interquartile range (Q1--Q3); horizontal line: median; whiskers: 1st--99th percentile. ($\diamond$): $\Gamma/r_1(V_\mathrm{med}/n_c)$; ($\triangle$): $\Gamma(1/r_1-1/r_2)$, both evaluated at the effective volume per meniscus $V_\mathrm{eff} = V_\mathrm{med}/n_c$ ($n_c = n_\mathrm{grains}-1$).}
  \label{fig:uc_distributions}
\end{figure}
Following Lian \emph{et al.}~\cite{Lian1993} (Eqs.~[16]--[17]), the two principal radii of curvature of the toroidal meniscus are
\begin{align}
\rho^*_1 &= \frac{1-\cos\phi}{\cos(\phi+\theta)}, \label{eq:rho1}\\[4pt]
\rho^*_2 &= \sin\phi - \frac{[1-\sin(\phi+\theta)](1-\cos\phi)}{\cos(\phi+\theta)},
\label{eq:rho2}
\end{align}
where $R = d/2$ is the grain radius, $\rho^*_i = r_i/R$, $r_1$ is the neck (concave) radius, and $r_2$ is the meridional (convex) radius of curvature. The volume of the toroidal liquid bridge is
\begin{equation}
V(\phi) = 2\pi^2 r_1 r_2^2 = 2\pi^2 R^3\,\rho^*_1\,(\rho^*_2)^2,
\label{eq:Vphi}
\end{equation}
which is the standard volume formula for an anchor ring (torus) of tube radius $r_1$ and centerline radius $r_2$. The capillary pressure follows from the Young--Laplace equation,
\begin{equation}
u_c = \Gamma\!\left(\frac{1}{r_1} - \frac{1}{r_2}\right)
    = \frac{\Gamma}{R}\!\left(\frac{1}{\rho^*_1} - \frac{1}{\rho^*_2}\right),
\label{eq:uc}
\end{equation}
where $\Gamma$ is the gas--liquid surface tension. (The factor $\cos\theta$ is omitted since $\theta \simeq 5^\circ$ and $\cos\theta \simeq 0.996 \approx 1$.)

For each liquid domain of measured volume $V_w$, the filling angle $\phi$ is determined by solving $V(\phi) = V_w$ numerically using the bisection method on the interval $\phi \in (0.02,\,\pi/2 - \theta)$. The function $V(\phi)$ is strictly monotonic on this interval (both $\rho^*_1$ and $\rho^*_2$ increase monotonically with $\phi$ for $\theta < 40^\circ$), guaranteeing a unique solution. The capillary pressure $u_c$ is then evaluated from Eq.~\eqref{eq:uc}.
 
The toroidal model is preferred over direct curvature measurement from the Marching Cubes mesh because the 8\,$\mu$m voxel resolution is insufficient to resolve the neck radius $r_1 \simeq 2$--$50\,\mu$m of typical pendular bridges ($r_1/\delta_\mathrm{vox} \simeq 0.3$--$6$). A sphere-of-radius test confirms that the cotangent Laplacian estimator overestimates the mean curvature by a factor of $\sim 6$ at this resolution, while the toroidal inversion is resolution-independent and relies only on the robustly measured liquid volume $V_w$.
 
The inversion was validated against the exact numerical solution of the Laplace--Young equation tabulated in Ref.~\cite{Lian1993} for a range of dimensionless volumes $V^* = V/R^3 \in [10^{-4},\,0.1]$ and contact angles $\theta \in [0^\circ,\,10^\circ]$. The relative error on $u_c$ is below 5\% for all stable bridge configurations encountered experimentally ($V_w \in [10^3,\,10^7]\,\mu\mathrm{m}^3$, $\phi \lesssim 35^\circ$).
 
For liquid domains connecting more than two grains (dimers, trimers, and larger clusters), the single-bridge inversion $V(\phi) = V_w$ is not directly applicable, since the total volume $V_w$ is distributed among $n_c$ individual menisci, where $n_c$ denotes the number of wetted grain--grain contacts within the domain. Because $n_c$ cannot be determined directly from the watershed segmentation alone --- which only provides the number of adjacent grain labels $n_g$ --- we approximate $n_c \approx n_g - 1$, corresponding to a minimum-connectivity (chain or spanning-tree) topology. This estimate is exact for dimers ($n_g = 3$, $n_c = 2$) and linear trimers ($n_g = 4$, $n_c = 3$), and constitutes a lower bound for more compact clusters (e.g.\ a tetrahedral cluster has at most $\binom{n_g}{2}$ contacts). An effective per-meniscus volume $V_\mathrm{eff} = V_w/n_c$ is then inverted through Eq.~\eqref{eq:Vphi} to yield $\phi$ and $u_c$. As shown in Fig.~\ref{fig:uc_distributions}, the median $u_c$ for dimers and trimers aligns with the theoretical prediction $\Gamma(1/r_1-1/r_2)$ evaluated at $V_\mathrm{eff}$ (triangle markers), validating the per-meniscus decomposition. Since multi-grain morphologies contribute at most $\sim 17\%$ of $P^\mathrm{micro}_\mathrm{cap}$ at $\epsilon = 0.05$, and systematically less at lower liquid content, the uncertainty introduced by the chain-topology approximation has a negligible effect on the macroscopic closure.

% ============================================================
%  BIBLIOGRAPHY
% ============================================================


\begin{thebibliography}{99}

\bibitem{Mitarai2006}
N. Mitarai and F. Nori,
Adv.\ Phys.\ \textbf{55}, 1 (2006).

\bibitem{Mani2013}
R. Mani, D. Kadau, and H.~J. Herrmann,
Granul.\ Matter \textbf{15}, 447 (2013).

\bibitem{Singh2014}
A. Singh, V. Magnanimo, K. Saitoh, and S. Luding,
Phys.\ Rev.\ E \textbf{90}, 022202 (2014).

\bibitem{Khamseh2015}
S. Khamseh, J.-N. Roux, and F. Chevoir,
Phys.\ Rev.\ E \textbf{92}, 022201 (2015).

\bibitem{Berger2016}
N. Berger, E. Az\'ema, J.-F. Douce, and F. Radjai,
Europhys.\ Lett.\ \textbf{112}, 64004 (2016).

\bibitem{Vo2020}
T.~T. Vo, S. Nezamabadi, P. Mutabaruka, J.-Y. Delenne, and F. Radjai,
Nat.\ Commun.\ \textbf{11}, 1 (2020).

\bibitem{Awdi2023}
A. Awdi, C. Chateau, F. Chevoir, J.-N. Roux, and A. Fall,
J.\ Rheol.\ \textbf{67}, 365 (2023).

\bibitem{Badetti2018a}
M. Badetti, A. Fall, D. Hautemayou, F. Chevoir, P. Aimedieu, S. Rodts, and J.-N. Roux,
J.\ Rheol.\ \textbf{62}, 1175 (2018).

\bibitem{Badetti2018b}
M. Badetti, A. Fall, F. Chevoir, and J.-N. Roux,
Eur.\ Phys.\ J.\ E \textbf{41}, 1 (2018).

\bibitem{Amarsid2024}
L. Amarsid, A. Awdi, A. Fall, J.-N. Roux, and F. Chevoir,
J.\ Rheol.\ \textbf{68}, 523 (2024).

\bibitem{Willett2000}
C.~D. Willett, M.~J. Adams, S.~A. Johnson, and J.~P. Seville,
Langmuir \textbf{16}, 9396 (2000).

\bibitem{Huang2015}
H. Huang, M. Sukop, and X. Lu,
\textit{Multiphase Lattice Boltzmann Methods: Theory and Application}
(John Wiley \& Sons, 2015).

\bibitem{Lian1993}
G. Lian, C. Thornton, and M.~J. Adams,
J.\ Colloid Interface Sci.\ \textbf{161}, 138 (1993).

\bibitem{Delenne2015}
J.-Y. Delenne, V. Richefeu, and F. Radjai,
J.\ Fluid Mech.\ \textbf{762}, R5 (2015).

\bibitem{Younes2023}
N. Younes, A. Wautier, R. Wan, O. Millet, F. Nicot, and R. Bouchard,
Comput.\ Geotech.\ \textbf{162}, 105677 (2023).

\bibitem{Richefeu2016}
V. Richefeu, F. Radjai, and J.-Y. Delenne,
Comput.\ Geotech.\ \textbf{80}, 353 (2016).

\bibitem{Mandal2020}
S. Mandal, M. Nicolas, and O. Pouliquen,
Proc.\ Natl.\ Acad.\ Sci.\ U.S.A.\ \textbf{117}, 8366 (2020).

\bibitem{Gans2020}
A. Gans, O. Pouliquen, and M. Nicolas,
Phys.\ Rev.\ E \textbf{101}, 032904 (2020).

\bibitem{Pouliquen2025}
O. Pouliquen,
Rheol.\ Acta (2025).

\bibitem{Iveson2001}
S.~M. Iveson, J.~D. Litster, K. Hapgood, and B.~J. Ennis,
Powder Technol.\ \textbf{117}, 3 (2001).

\bibitem{Awdi2025}
A. Awdi, C. Chateau, A. Fall, J.-N. Roux, and P. Aimedieu,
Granul.\ Matter \textbf{27}, 1 (2025).

\bibitem{Fisher1926}
R. Fisher,
J.\ Agric.\ Sci.\ \textbf{16}, 492 (1926).

\bibitem{Rumpf1974}
H. Rumpf and H. Schubert,
J.\ Chem.\ Eng.\ Jpn.\ \textbf{7}, 294 (1974).

\bibitem{Duriez2018}
J. Duriez and R. Wan,
Geomech.\ Energy Environ.\ \textbf{15}, 10 (2018).

\bibitem{Dragonfly2022}
Dragonfly 2022.1 [computer software] (2022),
\url{https://dragonfly.comet.tech}.

\bibitem{Liu2015}
T. Liu, A. Merat, M.~H.~M. Makhmalbaf, C. Fajardo, and P. Merati,
Exp.\ Fluids \textbf{56}, 166 (2015).

\bibitem{Fiscina2012}
J. Fiscina, M. Pakpour, A. Fall, N. Vandewalle, C. Wagner, and D. Bonn,
Phys.\ Rev.\ E \textbf{86}, 020103 (2012).

\bibitem{Schott2025}
F. Schott, B. Dollet, S. Santucci, C.~M. Schlep\"utz, C. Claudet,
S. Gst\"ohl, C. Raufaste, and R. Mokso,
Nat.\ Commun.\ \textbf{16}, 9210 (2025).

\end{thebibliography}
\end{document}